\documentclass[twocolumn, tightenlines, amsmath, amssymb, superscriptaddress, table,xcdraw]{revtex4-1}
\usepackage{graphicx} 
\usepackage{fourier}
\usepackage{physics}
\usepackage[separate-uncertainty=true]{siunitx}
\usepackage{hyperref}
\usepackage{amsmath}  
\usepackage{amsthm} 
\usepackage{dsfont}
\usepackage{booktabs} 
\usepackage{changes}
\usepackage[]{xcolor}

\newtheorem*{prop}{Proposition}

\newcommand{\hvec}[1]{\hat{\vec{#1}}}  

\begin{document}

\title{
Scalable machine learning-assisted clear-box characterization for optimally controlled photonic circuits
}

\author{Andreas Fyrillas}
\affiliation{Quandela, 7 Rue Léonard de Vinci, 91300 Massy, France}
\affiliation{Centre for Nanosciences and Nanotechnologies, CNRS, Université Paris-Saclay, UMR 9001, 10 Boulevard Thomas Gobert, 91120, Palaiseau, France}
\author{Olivier Faure}
\author{Nicolas Maring}
\author{Jean Senellart}
\affiliation{Quandela, 7 Rue Léonard de Vinci, 91300 Massy, France}
\author{Nadia Belabas}
\affiliation{Centre for Nanosciences and Nanotechnologies, CNRS, Université Paris-Saclay, UMR 9001, 10 Boulevard Thomas Gobert, 91120, Palaiseau, France}

\begin{abstract}
\hrule
\vspace{0.5cm}
    Photonic integrated circuits offer a compact and stable platform for generating, manipulating, and detecting light. They are instrumental for classical and quantum  applications. Imperfections stemming from fabrication constraints, tolerances and operation wavelength impose limitations on the accuracy and thus utility of current photonic integrated devices. Mitigating these imperfections typically necessitates a model of the underlying physical structure and the estimation of parameters that are challenging to access. Direct solutions are currently lacking for mesh configurations extending beyond trivial cases.
    We introduce a scalable and innovative method to characterize photonic chips through an iterative machine learning-assisted procedure. Our method is based on a clear-box approach that harnesses a fully modeled virtual replica of the photonic chip to characterize. The process is sample-efficient and can be carried out with a continuous-wave laser and powermeters. The model estimates individual passive phases, crosstalk, beamsplitter reflectivity values and relative input/output losses. Building upon the accurate characterization results, we mitigate imperfections to enable enhanced control over the device.
    We validate our characterization and imperfection mitigation methods on a 12-mode Clements-interferometer equipped with 126 phase shifters, achieving beyond state-of-the-art chip control with an average \SI{99.77}{\%} amplitude fidelity on 100 implemented Haar-random unitary matrices.
\vspace{0.5cm}
\hrule
\end{abstract}

\maketitle

Photonic integrated circuits (PICs) incorporate optical components on a compact substrate, enabling the generation, manipulation, and detection of light \cite{Wang2020}. PICs have emerged as a compelling and versatile platform to manipulate light, thanks to an unprecedented stability, compactness and capability for scaling up. These miniature devices have showcased their potential to revolutionize photonic quantum computing \cite{Zhong2020, Madsen2022}, quantum communication \cite{Luo2023}, quantum cryptography \cite{Fyrillas2023} and quantum sensing \cite{Polino2019}. Beyond the quantum realm, PICs find utility in classical domains such as microwave photonics, optical beamforming and high-precision sensing \cite{Bogaerts2020}. Furthermore, PICs can be used to natively perform matrix-vector multiplications, offering the potential to propel the field of artificial intelligence forward \cite{Zhou2022}. 

PICs are in particular widely used to perform linear operations on light, featuring components such as static beamsplitters and tunable phase shifters which are controlled by voltages or electric currents. Nevertheless, these elements exhibit imperfections in current photonic devices that cannot be overlooked (see Fig.\ \ref{fig:virtual_replica}a). In terms of practical applications, these defects lead for instance to a severe degradation in performance of optical neural networks \cite{Banerjee2023} and a marked decrease in the fidelity of quantum gates \cite{Mower2015}. Optimal control of PICs despite deviation from ideal devices is thus a primary challenge for quantum and classical applications.

Self-configuration protocols for PICs exist and mitigate imperfections without requiring detailed knowledge about the device. PIC self-configuration is a viable option in specific use cases, requiring for instance to route light from one input to one output \cite{Pérez2020, Kondratyev2023}, or when dealing with particular photonic circuits \cite{Hamerly2022}. Otherwise, the light amplitude and phase transformation implemented by a PIC must be measured \cite{Laing2012, Rahimi-Keshari2013, Dhand2016, Spagnolo2017} and the phase shifters reconfigured in a trial-and-error approach until the targeted transformation is reached. This is however a costly and experimentally cumbersome operation hindering taking full advantage of PIC reconfigurability. Self-configuration may additionally break down in the presence of inhomogeneous output losses and crosstalk between phase shifters.

A promising strategy is to leverage the capabilities of machine learning. Neural networks have been successfully trained in a black-box approach to connect the measured single-photon statistics produced by a 3-mode PIC to the voltages applied on 2 phase shifters \cite{Cimini2021}. Neural networks were also used in intermediate gray-box approaches where the algorithm only learns the Hamiltonian of the photonic device and the measurement probabilities are computed according to the laws of quantum mechanics \cite{Youssry2023}. In both cases, scalability is a major issue as the number of required data samples to train the neural network grows heavily with the complexity of the physical PIC.

In the light of self-configuration and neural network-based approaches, an ideal method for achieving PIC optimal control should possess the dual characteristics of adaptability to address various defect types and scalability concerning the number of components within a PIC. To that end, we adopt here a clear-box approach relying on a model of the PIC and its imperfections derived from physical intuition. In clear-box methods, the model is constrained to learn only the parameters of interest which is a promise for enhanced sample efficiency. However, the efficiency of this transparent approach hinges on the precision and faithfulness of the modeling of imperfections present in the physical system.
Operating in the clear-box paradigm implies that each imperfection type must be addressed with a tailored mitigation strategy. Errors in beamsplitter reflectivity, for instance, can be compensated by computing rectified phase shifts either optimized globally via gradient-based methods \cite{Mower2015, Burgwal2017} or optimized locally with faster deterministic schemes \cite{Bandyopadhyay2021, Kumar2021}. Similarly, compensation of crosstalk between phase shifters can in theory be achieved for optical neural networks \cite{Zhu2020}, and has been demonstrated in simple experimental cases \cite{Metcalf2014}. 
For clear-box imperfection mitigation, an accurate prior modeling and characterization of the PIC imperfections is essential. However, accessing directly the values of individual model parameters may be very challenging depending on the specific PIC architecture. This is especially true for universal-scheme PICs (Reck \cite{Reck1994}, Clements \cite{Clements2016}, Bell \cite{Bell2021}) which are notoriously hard to characterize, yet constitute the backbone of near-term photonic quantum processors \cite{Maring2023}.

We present an iterative machine learning-assisted PIC characterization process. We harness the sample-efficiency of the clear-box approach which paves the way for scalability in the characterization of increasingly bigger PIC architectures. We also exploit the large data processing abilities of machine learning to handle the large resulting number of physically meaningful parameters and complex interferometer meshes. A comparable characterization strategy has recently been mentioned in \cite{Bandyopadhyay2022}, with limited elaboration on the methodology. Our method offers valuable insights into the physics of the device which can then be used to improve fabrication processes, in contrast to neural network models that lack interpretability. In addition, we require only a laser and powermeters, or alternatively a single-photon source and single-photon detectors. 
The results of the characterization process are subsequently harnessed by a custom imperfection mitigation. We achieved  unparalleled optimal control on a 12-mode Clements universal interferometer, one of the largest PICs in terms of number of components currently available (see state of the art in App.\ \ref{app:soa}).

\begin{itemize}
    \item In section \ref{sec:modelling}, we model the physical linear PIC to characterize and the relevant imperfections to take into account. 
    \item Section \ref{sec:protocol} presents the different stages of the characterization protocol, that allow to finetune the modelled imperfections parameters. The protocol is then simulation benchmarked to demonstrate its effectiveness to converge to the true parameter values.
    \item Harnessing the knowledge gained by the characterization step, Section \ref{sec:mitigation} details our imperfection mitigation that translates targeted unitary matrices or sets of targeted phase shifts into voltages/electric currents and implements the target with high fidelity on the PIC.
    \item In Section \ref{sec:ascella}, we experimentally validate our characterization process on a 12-mode Clements-interferometer PIC featuring 126 thermo-optic phase shifters and 132 directional couplers. We characterize passive phases, thermal crosstalk, beamsplitter reflectivity errors and relative input/output losses. Using our imperfection mitigation, we implement unitary operations on single-photons and demonstrate beyond state-of-the-art 99.77 \% fidelity to the target.
\end{itemize}

We focus in the following on PICs for linear optics featuring beamsplitters and phase shifters, but our approach is generalizable to PICs manipulating for instance polarization of photons and featuring non-linear optical elements.

\section{Modelling photonic integrated circuit imperfections}
\label{sec:modelling}

\begin{figure*}
    \centering
    \includegraphics[width=\textwidth]{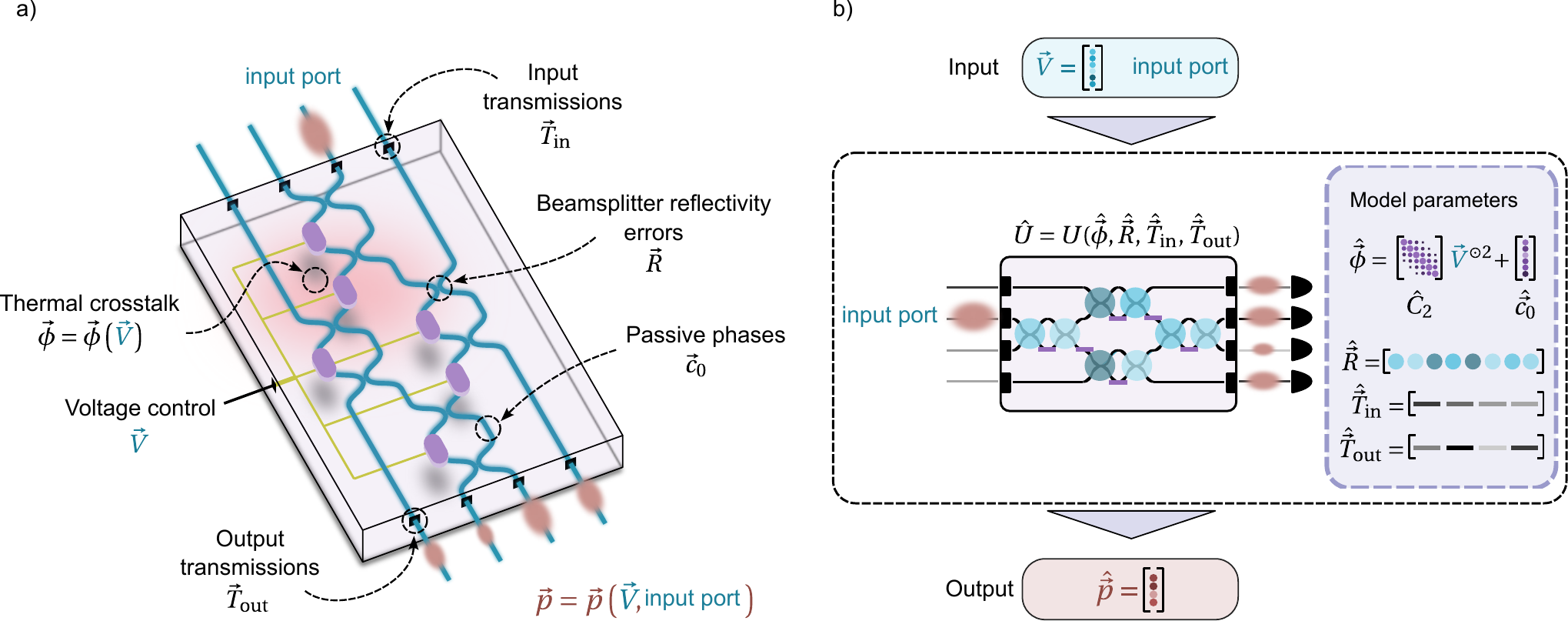}
    \caption{
    \textbf{Photonic chip imperfections modeled in a virtual replica.}  
    \textbf{a)} Physical photonic integrated circuits (PICs) often exhibit various imperfections resulting from fabrication constraints, tolerances and operation wavelength, illustrated here on a simplified PIC. In general, input and output ports have different optical transmissions, stored in vectors $\vec{T}_\text{in}$ and $\vec{T}_\text{out}$. In addition, the real beamsplitter reflectivity values $\vec{R}$ deviate from the target. Phase shifters (purple components) dissipating heat entail a phase-voltage relation of the type $\vec{\phi} = \vec{\phi}(\vec{V})$ between all the physical phase shifts $\vec{\phi}$ and applied voltages $\vec{V}$. In addition, optical path variations lead to non-zero phase shifts even without any voltages applied, i.e.\ $\vec{\phi}(\vec{0})=\vec{c}_0 \neq \vec{0}$. When sending light into the PIC, here represented by a laser pulse, the output light intensity distribution $\vec{p}$ depends on the applied voltages and the chosen input port.
    \textbf{b)} Our characterization process uses a virtual replica of the physical PIC. Hardware imperfections are modeled in the replica following Section \ref{sec:modelling}. The model parameters represent the replica current knowledge of the physical PIC characteristics: matrix phase-voltage relation $\hvec{\phi} = \hat{C}_2 \cdot \vec{V}^{\odot 2} + \hvec{c}_0$ (see Eq.\ \ref{eq:phase-voltage}), optical input/output transmissions $\hvec{T}_\text{in}$ and $\hvec{T}_\text{out}$ and beamsplitter reflectivities $\hvec{R}$, where the hat notation indicates predicted quantities. 
    When given a list of voltages $\vec{V}$, the model predicts the implemented phases $\hvec{\phi}$ on the virtual PIC and generates the matrix  $\hat{U} = U(\hvec{\phi}, \hvec{R}, \hvec{T}_\text{in}, \hvec{T}_\text{out})$ that encapsulates the virtual PIC action on light. The model then computes the predicted output light intensity distribution $\hvec{p}$, normalized such that its elements sum to 1, resulting from light injected into a single input port.
    }
    \label{fig:virtual_replica}
\end{figure*}

The relevant imperfections in PICs for linear optics are as shown on Fig. \ref{fig:virtual_replica}a:

\begin{itemize}
    \item Beamsplitter reflectivity deviating from the target value. On-chip beamsplitters are realized by directional couplers \cite{Marcatili1969}. Fabrication introduces random and systematic errors. Systematic errors also occur due to deviations of the light wavelength of the user from the fabrication wavelength. For universal-scheme PICs, which can in practice implement any unitary matrix acting on the spatial input modes, beamsplitter errors reduce the number of implementable unitary matrices \cite{Burgwal2017, Russell2017}.
    \item Passive phases \cite{Yang2015} due to waveguide length differences and inhomogeneities of the refractive index. As a consequence, phase shifters induce a non-zero phase even when no voltage/electric current is applied. Passive phases add a layer of difficulty to the characterization process.
    \item Crosstalk induced by reconfigurable components. For instance, heat produced by a thermo-optic phase shifter \cite{Harris2014} diffuses and distorts the action of other phase shifters \cite{Milanizadeh2019}. Crosstalk is also expected in the case of phase shifters harnessing strain-induced birefringence \cite{Dong2022} or the electro-optic effect \cite{Li2020}. Electric crosstalk can occur if the phase shifters share a common electric ground.
    \item Inhomogeneous input and output optical transmissions, due to differences in the optical coupling to the PIC or in light detection efficiencies. In practice, absorption losses in the PIC waveguides may not be entirely homogeneous. Nevertheless, we will assume uniformity of internal absorption losses.
\end{itemize}

The number of modes $m$ of a PIC is the number of input/output ports. The physical action of a PIC is encapsulated in an $m \times m$ matrix $U$ whose element $|u_{i,j}|^2$ represents the probability that photons injected in the input port $j$ exit through the output port $i$. This picture is valid both in classical electrodynamics and in quantum optics. $U$ may be non unitary to account for losses in the system. We use the following convention for an on-chip beamsplitter of reflectivity $R$:

\begin{equation}
    \begin{bmatrix}
        \sqrt{R}    & i\sqrt{1-R} \\
        i\sqrt{1-R} & \sqrt{R} 
    \end{bmatrix}.
\end{equation}
The matrix $U(\vec{\phi}, \vec{R})$ implemented by a PIC with phases $\vec{\phi}$ on its phase shifters and beamsplitters of reflectivity $\vec{R}$ is the matrix product of its individual components. If the $i^\text{th}$ input (resp. output) port has optical transmission $T$, we model this by multiplying the $i^\text{th}$ column (resp. row) of $U(\vec{\phi}, \vec{R})$ by $\sqrt{T}$. The lists of optical input and output transmissions are denoted $\vec{T}_\text{in}$ and $\vec{T}_\text{out}$. We normalize the maximum value of $\vec{T}_\text{out}$ to 1, hence $\vec{T}_\text{in}$ contains both the input inhomogeneities and the uniform global setup losses. The resulting $m\times m$ matrix modelling the PIC is denoted $U(\vec{\phi}, \vec{R}, \vec{T}_\text{in}, \vec{T}_\text{out})$. 

In the following we consider that the PIC phase shifters are voltage-controlled, but the case of electric current control is treated in an analog way. To model crosstalk between phase shifters, we use a phase-voltage relation linking the physically implemented phases $\vec{\phi}$ and the applied voltages $\vec{V}$ by a matrix relation of the type

\begin{equation}
\label{eq:phase-voltage}
    \vec{\phi} = \sum_{k\geq1} C_k \cdot \vec{V}^{\odot k} + \vec{c}_0
\end{equation}
where $\vec{c_0}$ is a vector with $n_\text{PS}$ entries containing the passive phases ($n_\text{PS}$ is the number of on-chip phase shifters), $^{\odot}$ represents element-wise exponentiation, and the $C_k$ are $n_\text{PS} \times n_\text{PS}$ matrices. Off-diagonal elements in the $C_k$ account for crosstalk between phase shifters. In principle it is sufficient for thermo-optic phase shifters to keep only the passive phases and the $\vec{V}^{\odot 2}$ term, because of the $V^2$-dependence of Joule heating. Optionally, adding a $\vec{V}^{\odot 4}$ term takes into account the change of heater resistance with temperature. We discuss in App.\ \ref{app:electric_crosstalk} the case of electric crosstalk that is possibly present in PICs with voltage-controlled phase shifters. In the following, we will use without loss of generality a phase-voltage relation of the form $\vec{\phi} = C_2 \cdot \vec{V}^{\odot 2} + \vec{c}_0$. 

\section{Iterative machine learning-assisted PIC characterization}
\label{sec:protocol}

\subsection{Virtual PIC replica}
\label{sec:replica}

Our method relies on a virtual replica model of the physical PIC to characterize. The replica depicted on Fig.\ \ref{fig:virtual_replica}b is endowed with imperfections modeled as described in Section \ref{sec:modelling}. The model parameters are the matrix $\hat{C}_2$ and the passive phase vector $\hvec{c}_0$ included in the phase-voltage relation Eq. \ref{eq:phase-voltage}, the optical input/output transmission vectors $\hvec{T}_\text{in}$ and $\hvec{T}_\text{out}$ and the beamsplitter reflectivities $\hvec{R}$. The hat notation indicates predicted quantities.

The virtual replica predicts the output light intensity distribution $\hvec{p}(\vec{V},i)$ expected at the output of the physical PIC when light is injected in input $i$ and voltages $\vec{V}$ are applied. To do so, the replica uses the learned model parameters to compute the predicted phase shifts $\hvec{\phi}$ and construct the matrix $\hat{U} = U(\hvec{\phi}, \hvec{R}, \hvec{T}_\text{in}, \hvec{T}_\text{out})$ describing the replica action on light. The resulting output light intensity distribution $\hvec{p}$ is computed from $\hat{U}$ and the input port index and normalized afterwards to sum to 1. 

The model parameters are optimized along our characterization process with the aim that the difference between the measured output light intensity distributions $\vec{p}(\vec{V},i)$ and the predicted $\hvec{p}(\vec{V},i)$ is minimized. The replica is initialized as in Fig.\ \ref{fig:process}a: $\hat{C}_2$ is the zero matrix, $\vec{c}_0=\vec{0}$, beamsplitters have their target reflectivity value and $\hvec{T}_\text{in}^{(i)}=\hvec{T}_\text{out}^{(i)}=1$.

\subsection{Voltage interference fringe measurement}
\label{sec:v_ifm}

\begin{figure*}[ht]
    \centering
    \includegraphics[width=\textwidth]{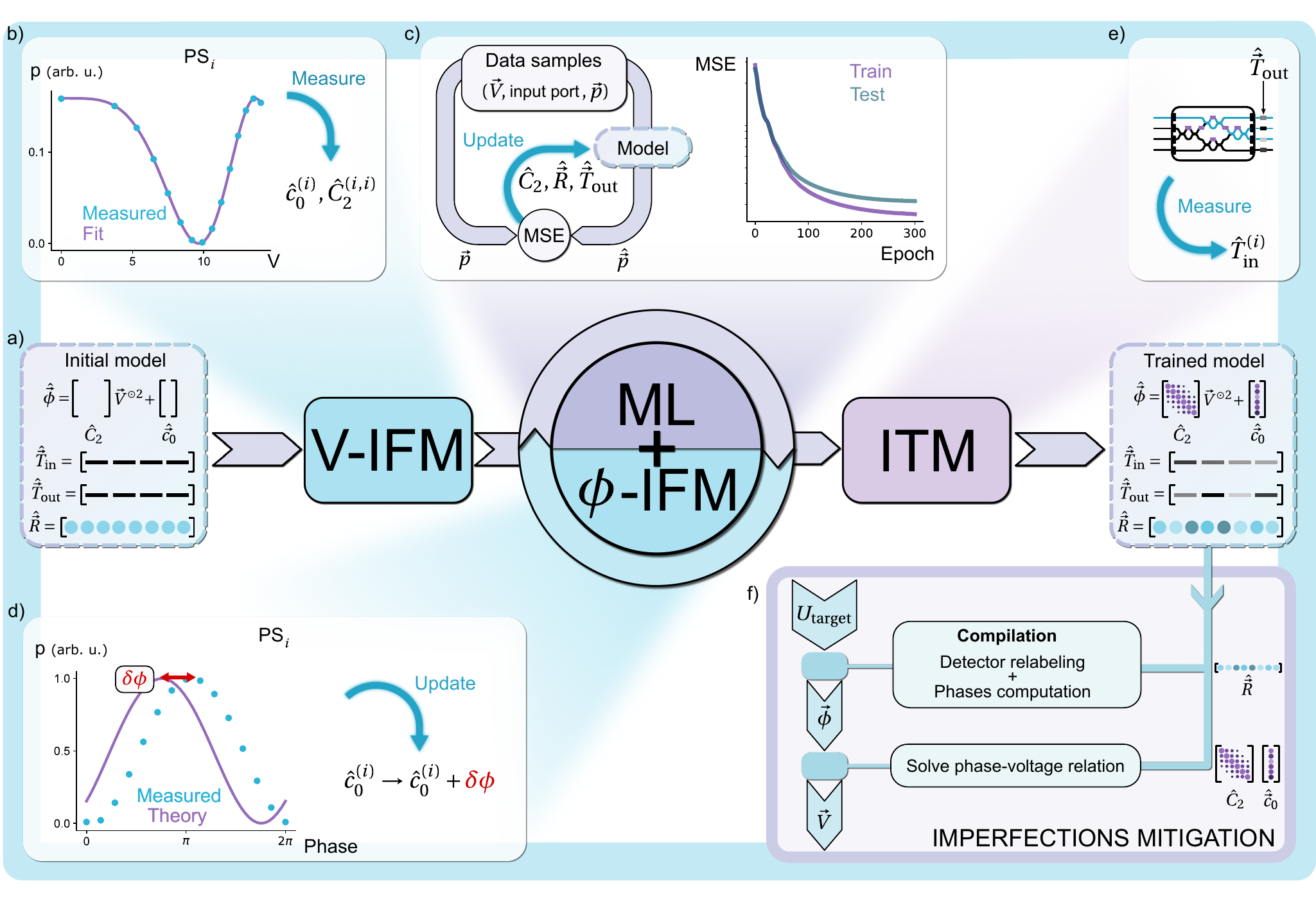}
    \caption{
    \textbf{Characterization of photonic integrated circuits using an iterative machine learning-assisted process and harnessed by an imperfection mitigation.}
     \textbf{a)} In our photonic integrated circuit (PIC) characterization process, the virtual replica model is initialized with parameter values given in Sec.\ \ref{sec:replica} that are optimized subsequently.
     \textbf{b)} The first step in the characterization process is a voltage interference fringe measurement (V-IFM), detailed in Sec.\ \ref{sec:v_ifm}. Each on-chip phase shifter is individually swept in voltage. Fitting each recorded interference fringe allows to populate the passive phases vector $\hvec{c}_0$ and diagonal elements of the matrix $\hat{C}_2$ in the phase-voltage relation.
     \textbf{c)} The model is subsequently fine-tuned by a machine learning (ML) step. The ML step requires a dataset of the form $\{(\vec{V}, \text{input port}, \vec{p})\}$ acquired as described in Sec.\ \ref{sec:ml}. ML consists in a gradient-descent algorithm that updates the $\hat{C}_2$ matrix, the beamsplitter reflectivity vector $\hvec{R}$ and output transmissions $\hvec{T}_\text{out}$. The minimized cost function is the mean square error (MSE) between the data sample output light intensities $\vec{p}$ and corresponding model predictions $\hvec{p}$. 
     \textbf{d)} Phase interference fringe measurement ($\phi$-IFM): the learned phase-voltage relation is solved to sweep the phase of the individual phase shifters. The offset between the measured data points and the expected curve is used to update the passive phases $\hvec{c_0}$. The process does multiple ML + $\phi$-IFM iterations until the MSE stops improving compared to the previous iteration (see Sec.\ \ref{sec:p_ifm}). 
     \textbf{e)} The last step is an input transmission measurement (ITM). Light intensities are measured without normalization on the physical PIC from each used input. Differential output losses are compensated using the estimated model parameters, yielding a measurement of $\vec{T}_\text{in}$. Further information in Sec.\ \ref{sec:itm}.
     \textbf{f)} The parameters of the fully trained model are harnessed in our imperfection mitigation (see Sec.\ \ref{sec:mitigation}). To implement a target unitary matrix $U_\text{target}$ with the PIC, a compilation step first relabels the detector outputs and computes a set of phase shifts $\vec{\phi}$ that faithfully recreates the target $U_\text{target}$ taking into account the learned beamsplitter reflectivity values $\hvec{R}$ (see Sec.\ \ref{sec:compilation}). The phase shifts are then converted into voltages $\vec{V}$ by solving the learned phase-voltage relation  $\hvec{\phi}=\hat{C}_2 \cdot \vec{V}^{\odot2} + \hvec{c}_0$ (see Sec.\ref{sec:crosstalk}). $\vec{V}$ is afterwards communicated to the physical PIC. 
    }
    \label{fig:process}
\end{figure*}

The first stage denoted V-IFM in our characterization protocol illustrated on Fig.\ \ref{fig:process}b is an interference fringe measurement. V-IFM contributes to establishing the phase-voltage relation $\vec{\phi} = C_2 \cdot \vec{V}^{\odot 2} + \vec{c}_0$ of the PIC. For each phase shifter PS$_i$, light is injected into the PIC via a single input chosen such that light is routed to PS$_i$ using only already characterized phase shifters and then exits through a monitored output. The voltage applied on PS$_i$ is swept, starting from 0 V to up to a designated safe maximum. The voltage sweep produces an oscillating output light intensity that is recorded and fitted to recover initial values for PS$_i$'s self-heating coefficient $C_2^{(i,i)}$ and passive phase $c_0^{(i)}$. To accommodate various PIC mesh designs, we algorithmically generate our interference fringe measurement protocol. This protocol provides the phase shifter characterization sequence, the routing of light from input to output for measurement, and phase offsets that need to be accounted for (see App.\ \ref{app:charac_algo}). Most experimental PIC characterizations \cite{Qiang2018, Adcock2019, Arrazola2021} only carry out the V-IFM, but in the general case, the retrieved passive phase $c_0^{(i)}$ is then flawed because of the imperfect directional coupler reflectivities and uncompensated crosstalk between phase shifters (see App.\ \ref{app:ifm_imperfect}).

\subsection{Fine tuning the model parameters via machine learning}
\label{sec:ml}

Estimating physical PIC parameters like crosstalk between phase shifters is a challenging task due to the substantial number of parameters $n_\text{PS}^2$ that need to be determined, with $n_\text{PS}$ the number of phase shifters. Therefore, we employ machine learning (ML) to find optimized values for the model parameters $\hat{C}_2$, $\hvec{R}$ and $\hvec{T}_\text{out}$ as shown on Fig.\ \ref{fig:process}c.

Our method requires a number of training samples $n_\text{train}$ of the order of the number of parameters to optimize ($n^2_\text{PS}$ + number of beamsplitters + number of modes). This is discussed in Sec.\ \ref{sec:bench_charac}. For each data sample, a set of random voltages $\vec{V}$ is generated and applied on the phase shifters. Light is then injected into a randomly chosen input port, and the corresponding output light intensity distribution $ \vec{p}$ is measured and normalized to sum to 1. Depending on the PIC layout, gathering data samples only from a restricted number of input ports might be sufficient. The data samples are a collection of the form $\bigl\{ (\vec{V}, \text{port}, \vec{p}) \bigl\}$.

We also acquire a set of test $n_\text{test}$ samples with a ratio $n_\text{train}/n_\text{test} = 80/20$. The gradient-descent algorithm is implemented with the \texttt{PyTorch} package in \texttt{Python} \cite{Pytorch}. The Adam optimizer \cite{Adam} updates the model parameters via stochastic gradient descent to minimize the mean squared error (MSE) between the training set $\{ \vec{p} \}$ and the model predictions $\{ \hvec{p} \}$ for output light intensity distributions. The model is evaluated on the set of test samples to monitor the progress of the learning process. Each model parameter type is tuned with a different learning rate.

\subsection{Phase interference fringe measurements and iterating the process}
\label{sec:p_ifm}

The estimated passive phases $\hvec{c}_0$ remain fixed during the ML stage because the model usually converges to a local minimum for the passive phases. We introduce a step ($\phi$-IFM on Fig. \ref{fig:process}d) that updates $\hvec{c}_0$ while leveraging the knowledge gained by the virtual replica in the ML stage. The so-far learned phase-voltage relation is solved to sweep the phase of each phase shifter from 0 to $2\pi$ while compensating crosstalk. The passive phases $\vec{c}_0$ are then updated by estimating the offset between the measured and expected interference fringe, generated by taking into account the learned beamsplitter reflectivity values and output transmissions (details in App.\ \ref{app:ifm_principle}).

The automated characterization protocol performs multiple (ML + $\phi$-IFM) iterations, gradually acquiring more precise information about the physical device. The gradient descent learning rates of the virtual replica are all divided by a constant factor after each iteration to enable faster convergence to the optimum. The protocol exits the iteration loop when the minimum MSE measured on the test dataset during the ML stage gradient descent exceeds that of the previous iteration.

\subsection{Input transmission measurement}
\label{sec:itm}

The last stage in our characterization protocol measures the optical input transmissions of the PIC (ITM on Fig.\ \ref{fig:process}e). For each addressed input port $i$, we select among the training dataset the voltage configuration $\vec{V}$ yielding the best digital replica prediction. The voltage configuration is applied on the physical PIC and the output light intensity $\vec{p}$ is measured without normalization. Differential output transmissions are compensated by computing the vector $\vec{p} \oslash \hvec{T}_\text{out}$ where $\oslash$ is element-wise division and $\hvec{T}_\text{out}$ are the output transmissions estimated from the ML stages (see Fig.\ \ref{fig:process}c and Sec.\ \ref{sec:ml}). The sum of the components of this vector yields $T_\text{in}^{(i)} \times P$ where $T_\text{in}^{(i)}$ is the transmission of input port $i$ and $P$ is the input light intensity.

\subsection{Simulation benchmarking the PIC characterization method}
\label{sec:bench_charac}

\begin{figure*}
    \centering
    \includegraphics[width=\textwidth]{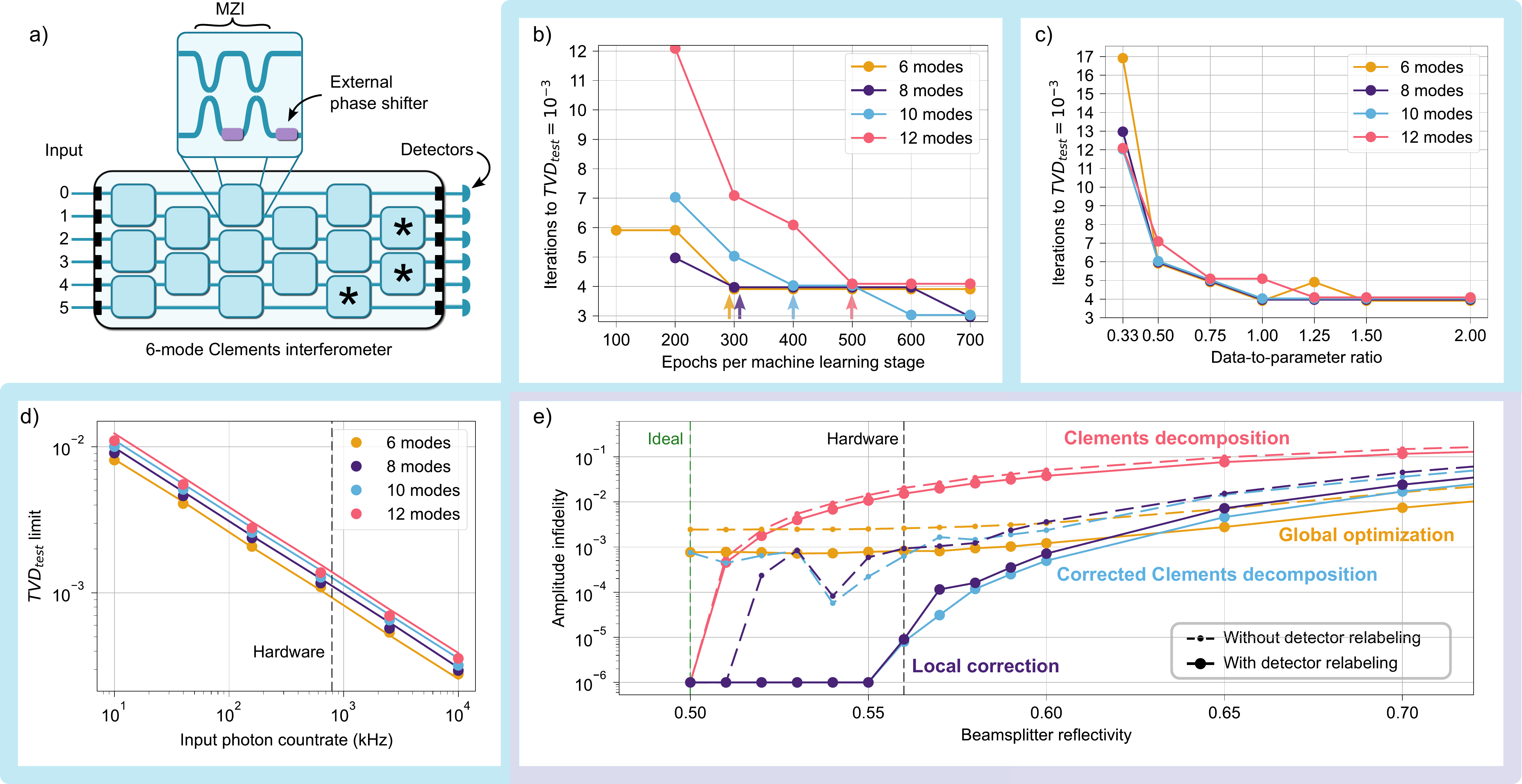}
    \caption{
    \textbf{Evaluating the photonic integrated circuit characterization method and the imperfection mitigation through simulation benchmarking}
     \textbf{a)} The characterization process and imperfection mitigation are simulation benchmarked on universal-scheme PICs with a Clements mesh with 6 (drawn here) up to 12 modes. Blue lines: waveguides. Blue square: mesh unit cell. The unit cell consists of a Mach-Zehnder interferometer (MZI) followed by an external phase shifter (see inset). Purple rectangles: reconfigurable phase shifter. Waveguides closely spaced form a beamsplitter. Unit cells marked with a star (*) do not feature an external phase shifters. b) to d): each point is a single simulation run.
     \textbf{b)} Impact of the number of gradient descent epochs per ML stage (see Sec. \ref{sec:ml}) on the number of required (ML + $\phi$-IFM) iterations to characterize a PIC. $\text{TVD}_\text{test}$ evaluates the accuracy of the model predictions (see App.\ \ref{app:pic_metrics}). 
     \textbf{c)} Number of (ML + $\phi$-IFM) iterations against the data-to-parameter ratio.
     \textbf{d)} Effect of photon counting noise on the learning process. Lowest reached $\text{TVD}_\text{test}$ as a function of the input single-photon countrate, assuming a detection integration time of 1 second with single-photon detectors. Continuous lines indicate the theoretical threshold. The black dotted line indicates the photon countrate for the experimental validation of our characterization process (see Sec.\ \ref{sec:ascella}). Details on the simulation benchmarking of the characterization protocol in Sec. \ref{sec:bench_charac}.
     \textbf{e)} Simulated comparison of compilation methods (see Sec.\ \ref{sec:compilation}). Average amplitude infidelity ($=1 - \text{amplitude fidelity}$, see App. \ref{app:amplitude_fidelity}) as a function of uniform beamsplitter reflectivity for a 12-mode Clements interferometer evaluated on 100 Haar-random unitary matrices. Dashed lines with small dots: standard method. Continuous line with big dots: method with prior detector relabelling. Red: Clements decomposition \cite{Clements2016}. Light blue: Clements decomposition with corrected unit cell \cite{Kumar2021}. Purple: Local deterministic phase correction \cite{Bandyopadhyay2021}. Yellow: Global phase optimization. Green dashed line: ideal reflectivity value for a Clements interferometer. Black dashed line: average reflectivity on our hardware (see Sec.\ \ref{sec:ascella}).  Value zero is clipped to $10^{-6}$.
    }
    \label{fig:benchmark_photograd}
\end{figure*}

We benchmark our characterization protocol on simulated universal-scheme PICs with a Clements mesh and 6 to 12 modes. (see Fig.\ \ref{fig:benchmark_photograd}a). The evaluation metric $\text{TVD}_\text{test}$ for the virtual replica prediction accuracy is the average total variation distance (TVD, see App.\ \ref{app:pic_metrics}) evaluated on the test dataset. $\text{TVD}_\text{test}$ is the average distance between the measured output light intensity distributions and the corresponding model predictions. 

\paragraph{Reduction of the characterization duration}
In practice, the time needed to characterize a PIC is predominantly consumed by the $\phi$-IFM stages, which are constrained in terms of speed by the PIC reconfiguration and light intensity integration times. Hence, reducing the number of (ML + $\phi$-IFM) iterations results in a significantly more time-efficient PIC characterization. We show on Fig.\ \ref{fig:benchmark_photograd}b that the number of (ML + $\phi$-IFM) iterations required to characterize a PIC can be substantially reduced by increasing the number of gradient descent epochs processed in the ML stage. For the following, we use for each PIC size the value indicated by the corresponding arrow on the plot. We note that the total number of epochs to process increases with the number of modes, as the precision required on model parameters increases with PIC complexity in order to minimize error propagation across an increasing number of component layers.

\paragraph{Sample efficiency}
We define the data-to-parameter ratio as the number training samples divided by the number of parameters trained during the ML stage.
Fig.\ \ref{fig:benchmark_photograd}c demonstrates that a data-to-parameter ratio of 1 is a good compromise between low iteration count, convergence guarantees and data acquisition time. This demonstrates that our clear-box PIC characterization is more sample efficient than black-box \cite{Cimini2021} and grey-box \cite{Youssry2023} alternatives (see App.\ \ref{app:soa}). We simulate in App.\ \ref{app:max_modes} the characterization of a 24-mode Clements interferometer and demonstrate that the sample efficiency also holds for significantly larger interferometer sizes.

\paragraph{Impact of photon-counting noise}
In practice, measurement noise naturally sets a limit on the minimally attainable $\text{TVD}_\text{test}$ denoted $\text{TVD}_\text{test,limit}$, due to the inherent difference between the noisy test dataset and the noiseless replica predictions. Knowing the order of magnitude of $\text{TVD}_\text{test,limit}$ is crucial from an experimental point of view to evaluate if the characterization protocol has converged to the physically allowed limit. We examine the case where single photons are injected into the PIC and detected by single-photon detectors. Simulated characterizations yield values $\text{TVD}_\text{test,limit}$ graphed as a function of the input photon countrate on Fig.\ \ref{fig:benchmark_photograd}d, assuming a detection integration time of 1 second. The reached $\text{TVD}_\text{test,limit}$ values agree with the theoretical threshold plotted as continuous lines. Estimating $\text{TVD}_\text{test,limit}$ in the case of laser light and powermeter arrays necessitates a case-by-case approach, as powermeter arrays commonly exhibit dominant Gaussian noise originating from internal components.

\section{Imperfection mitigation}
\label{sec:mitigation}

To mitigate imperfections, universal-scheme PICs benefit from a compilation process that translates target unitary matrices into phases to implement on the integrated phase shifters. To compute voltages from these phases, the PIC phase-voltage relation is retrieved from the trained virtual replica and solved. Imperfection mitigation is necessary to demonstrate enhanced PIC control (see Fig.\ \ref{fig:process}f), leveraging the estimated imperfections by our PIC characterization process.

\subsection{Compilation from unitary matrices to phases and comparison of methods}
\label{sec:compilation}

Universal-scheme PICs, like Clements interferometers (see Fig.\ \ref{fig:benchmark_photograd}a) with a tiled rectangular mesh of Mach-Zehnder interferometers (MZI), can by definition implement any unitary matrix acting on the spatial input modes. The compilation process computes the phase shifts that implement a target unitary matrix. Ideally, the compilation should deliver adequate phases leading to a high-fidelity reconstitution of the target matrix with minimal computation times to guarantee fast PIC reconfiguration times. While deterministic compilation methods achieve fast computation with a minimal overhead, another point to consider is that deterministic compilations are generally tied to a specific mesh e.g.\ Clements interferometers. In contrast, gradient descent based-methods work on any interferometer mesh. The optimized cost function of gradient-based methods may be in addition modified to compute phase shifts while adhering to additional specific constraints, for instance, to reduce the overall power consumption.

\paragraph{Comparison of compilation methods}
The "Clements decomposition" is the canonical deterministic compilation method for Clements interferometers \cite{Clements2016}, but assumes a uniform beamsplitter reflectivity value of 0.5. To cope with beamsplitter reflectivity errors, the "corrected Clements decomposition" \cite{Kumar2021} takes into account reflectivity errors while following the algorithm of the standard Clements decomposition. The "local correction" compiler \cite{Bandyopadhyay2021} adjusts deterministically the phases yielded by the Clements decomposition by looking at each Clements interferometer unit cell individually. The gradient based-compilation method, denoted "global optimization", optimizes all the phases globally to maximize the fidelity of the implemented matrix with respect to the target (see App.\ \ref{app:phase_optim}). 

We compare the different methods in simulations on 100 Haar-random target unitary matrices as shown on Fig.\ \ref{fig:benchmark_photograd}e (dashed lines). The simulation scenario is a 12-mode Clements interferometer with a uniform beamsplitter reflectivity error, in agreement with the features of our hardware (see Sec.\ \ref{sec:ascella}). This choice is also relevant for PICs in general because beamsplitters tend to exhibit spatially correlated reflectivity values \cite{Lu2017}. Consequently the effects of deterministic errors prevail over random fabrication errors in practice \cite{Bandyopadhyay2021, Banerjee2023}. Each target unitary matrix is compiled into phases taking into account the uniform reflectivity value. The computed phases are then applied on the simulated imperfect PIC. All simulations were carried out with the \texttt{Perceval} photonic circuit simulator \cite{Heurtel2023}. For each Haar-random matrix, the amplitude fidelity to the target (see App.\ \ref{app:amplitude_fidelity}) of the implemented matrix is computed. We observe that the Clements decomposition is strongly affected by beamsplitter reflectivity deviating from the ideal 0.5 value. Deterministic corrected methods "local correction" and "corrected Clements decomposition" achieve very similar good results, while "global optimization" shows an advantage for high reflectivity error.

\paragraph{Introducing prior detector relabeling}
We introduce in our compilation process a detector relabeling step before the phases computation, which significantly increases the fidelity of all of the presented methods (continuous lines on Fig.\ \ref{fig:benchmark_photograd}e). Intuitively, relabeling detectors for a 2-mode circuit is an easy imperfection mitigation technique. Indeed, when considering e.g.\ an MZI with uniform beamsplitter reflectivity values differing from 0.5, the \emph{cross} configuration is not feasible but the \emph{bar} configuration is (see App.\ \ref{app:mzi} for definitions). Permuting the labels of the two output detectors and setting the MZI to the \emph{bar} configuration, yields a perfect measured \emph{cross} configuration. On the full circuit level, obtaining the detector relabeling which maximizes the fidelity is less obvious. Iterating on all $m!$ possible permutations is unpractical, hence we randomly sample 32 random permutations, which yields a significant improvement as illustrated on Fig. \ref{fig:benchmark_photograd}e.
Detector relabelling has also been used in \cite{Kumar2021} to decrease PIC power consumption.

We discuss in App. \ref{app:losses_mitigation} a method for mitigating inhomogeneous input and output transmissions.

\subsection{Thermal crosstalk compensation}
\label{sec:crosstalk}

Knowing the phases $\vec{\phi}$ to apply on the PIC phase shifters, the phase-voltage matrix equation (Eq. \ref{eq:phase-voltage}) has to be solved for the voltages. Non-linear matrix equations are notoriously hard to solve analytically. We implement an iteration-based solver detailed in App.\ \ref{app:solver} to approximate a solution. The returned voltage configuration readily compensates thermal crosstalk because of the matrix formulation of the equation and implements the target phases within \SI{0.1}{\milli\radian} precision as set by our convergence threshold.

\section{Applying the process on a 12-mode universal-scheme PIC}
\label{sec:ascella}

\begin{figure*}[ht]
    \centering
    \includegraphics[width=\textwidth]{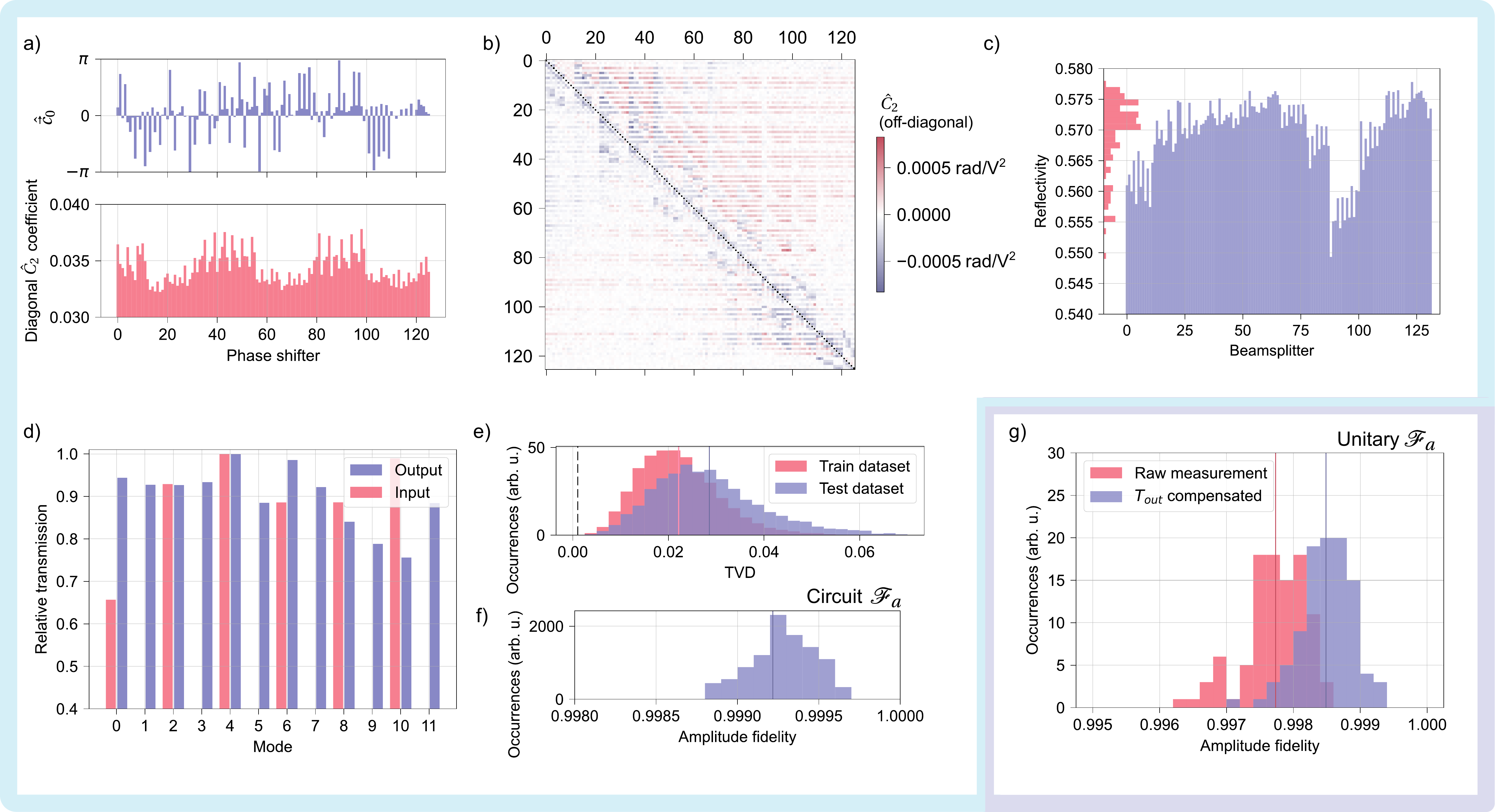}
    \caption{
    \textbf{Experimental validation of our photonic chip characterization protocol on a 12-mode universal-scheme interferometer.} 
    The phase-voltage relation of the PIC is of the form $\vec{\phi} = C_2 \cdot \vec{V}^{\odot 2} + \vec{c}_0$ (see Eq.\ \ref{eq:phase-voltage}). The top plot in \textbf{a)} depicts the estimated values of the passive phases vector $\hvec{c}_0$, and the bottom plot the diagonal elements of the matrix $\hat{C}_2$. 
    The off-diagonal elements of $\hat{C}_2$ are displayed in \textbf{b)}. They represent thermal crosstalk between phase shifters. 
    \textbf{c)} Reflectivity of each individual on-chip beamsplitter. Histogram of the values on the plot y-axis.
    \textbf{d)} The bar plot indicates the relative optical input and output transmissions of the PIC, normalized such that the maximum is 1 for each set. Notice that we only use 6 inputs of the PIC. 
    \textbf{e)} Histogram of total variation distance (TVD), that is the difference between the fully trained virtual replica predictions and the training/test dataset output light intensity distributions (see App. \ref{app:pic_metrics} for definition). Vertical lines indicate the average for the train dataset (\SI{2.2}{\%}) and for the test dataset (\SI{2.9}{\%}). 
    \textbf{f)} Histogram of measured circuit amplitude fidelity values for 100 random phase configurations implemented on the physical PIC (not Haar-random). For each phase configuration, the amplitude fidelity between the expected and the measured amplitude matrix is acquired. The expected matrix is generated by using the beamsplitter reflectivity values $\hvec{R}$ and output transmissions $\hvec{T}_\text{out}$ estimated by the virtual replica. The voltages to apply are computed by solving the learned phase-voltage relation. The vertical line indicates the average amplitude fidelity of \SI{99.93}{\%}.
    \textbf{g} Histogram of measured unitary amplitude fidelity values for 100 $12 \times 12$ Haar-random target unitary matrices. We post-process the measurement to compensate for output transmissions. Vertical lines indicate averages of \SI{99.77}{\%} and \SI{99.85}{\%}. See Sec.\ \ref{sec:ascella}.
    }
    \label{fig:validation}
\end{figure*}

We validate our process experimentally on a physical 12-mode universal-scheme PIC with a Clements mesh \cite{Taballione2021}. It features 126 reconfigurable thermo-optic phase shifters and 132 directional couplers. This is one of the biggest available PICs in terms of number of on-chip components (see Table \ref{tab:soa}). This showcases that our method scales beyond the few-phase-shifters case. We use a single-photon source, addressing only even input ports, and single-photon detectors on all output ports. It would be equivalent to use a continuous-wave laser and powermeters, but the measurement noise may be greater. We characterize the phase-voltage relation assumed to be of the form $\vec{\phi} = C_2 \cdot \vec{V}^{\odot 2} + \vec{c}_0$, beamsplitter reflectivity errors and input/output transmissions. The entire characterization process, including the data samples acquisition, took approximately 18 hours. See Fig.\ \ref{fig:validation}a-d for characterization results. In particular, the estimated $\hat{C}_2$ matrix on Fig.\ \ref{fig:validation}b has predominant values around its diagonal, meaning the estimated crosstalk is as expected weaker between distant components. The beamsplitter reflectivity values on Fig.\ \ref{fig:validation}c are strongly deviating from the ideal 0.5 value, because the single-photon source emission wavelength and the PIC fabrication wavelength are \SI{11}{\nano\meter} apart. 
The digital replica stabilizes at an average difference between its predictions and the test data samples $\text{TVD}_\text{test}\approx \SI{2.9}{\%}$, which is one order of magnitude higher than the expected value of $\SI{0.1}{\%}$ (see Fig.\ \ref{fig:benchmark_photograd}d). This discrepancy hints towards additional PIC imperfections not taken into account by the virtual replica.
We experimentally assess the accuracy of the characterization by implementing 100 random phase configurations $\vec{\phi}$ on the PIC phase shifters. For each configuration, the amplitude fidelity (see App.\ \ref{app:amplitude_fidelity}) is measured between the implemented matrix and the expected matrix, generated from the learned beamsplitter errors and output transmissions. We call this metric the circuit amplitude fidelity and measure $\mathcal{F}_a = \SI{99.92(2)}{\%}$ (see Fig.\ \ref{fig:validation}f), where the error bar is one standard deviation. In general, we empirically observe the relation $\mathcal{F}_a \approx 1 - \text{TVD}_\text{test}^2$ between the measured amplitude fidelity and $\text{TVD}_\text{test}$.

We then measure the average amplitude fidelity with respect to 100 $12\times 12$ Haar-random unitary matrices (see Fig.\ \ref{fig:validation}g). We compile the matrices into phases using the "local correction" method (see Sec.\ \ref{sec:compilation}) with detector relabeling based on the estimated beamsplitter reflectivity values $\hvec{R}$. We obtain a unitary amplitude fidelity $\mathcal{F}_a = \SI{99.77(4)}{\%}$, which represents the highest value recorded for Clements interferometers to this date (see Table \ref{tab:soa}). Correcting for the estimated output transmissions $\hvec{T}_\text{out}$, we arrive at a more precise evaluation of the amplitude fidelity of the actual implemented matrix on the PIC, with value $\mathcal{F}_a = \SI{99.85(4)}{\%}$. Both values are to our knowledge the highest reported in the literature, obtained on the most complex PIC characterized with machine learning so far (see Table \ref{tab:soa}).

\section{Discussion and outlook}

Our characterization method combines machine learning with a clear-box approach, modeling both the physical PIC to characterize and its imperfections. We thus overcome accuracy limitations imposed by  PIC characterization processes based solely on interference fringe measurements. Our method requires significantly lower amounts of training samples and computational power than neural network-based black- and grey-box methods (see Table \ref{tab:soa}) to train the model. Finally, our method does not depend on the interferometer structure and can therefore be used to handle complex interferometer meshes and large amounts of parameters, which ensures scalability

We validate our characterization and modeling method on a 12-mode Clements interferometer with a record value of \SI{0.08}{\%} amplitude infidelity between the measured PIC behavior and virtual replica model predictions.  

Furthermore, we have compared and enhanced imperfection mitigation methods. We demonstrate optimal chip control by combining imperfection mitigation with our PIC characterization protocol, allowing us to implement unitary matrices with \SI{0.23}{\%} amplitude infidelity, which is to our knowledge the best value in the literature. We emphasize that in addition we obtained it on one of the most complex available PICs.

Our process is straightforward  to implement experimentally and can be performed with only a continuous-wave laser and powermeters. In addition, our strategy can be tailored to model other on-chip components and imperfections, extending its applicability to various photonic circuits and systems. 
To reach even better PIC control, one route is to further investigate the modeling and characterization of other types of PIC imperfections. The development of faster compilation methods that account for and effectively mitigate these additional imperfections will also push the boundaries of PIC optimal control in the context of increasingly intricate and large interferometer meshes.

The clear-box approach finally guarantees transparency of the characterization process and holds the promise for improved fabrication processes by accurately probing individual  characteristics of photonic devices that are typically hard to reach, like thermal crosstalk or individual beamsplitter reflectivity values.

The increased reliability of photonic devices presents a transformative opportunity to rise to the current technological challenges  in telecommunications, data processing, sensing and quantum information processing. In particular, photonic quantum computing stands to reap substantial benefits from increased PIC accuracy, by achieving higher qubit fidelities. These advances  open the door to efficient near-term quantum processors with demonstrations of boson sampling with reconfigurable circuits and graph problem solvers and lay the foundations for fault-tolerant quantum computing harnessing integrated photonic components.

\section*{Methods}

\subsection*{Simulation benchmarking the characterization process}

We simulate Clements interferometers with 6, 8, 10 and 12 modes featuring a phase-voltage relation of the form $\vec{\phi}=C_2 \cdot \vec{V}^{\odot2} + \vec{c}_0$. The parameters of the simulated physical device that are to be estimated by the virtual replica are set as follows. The matrix $C_2$ is generated deterministically: the diagonal coefficients are set to 0.034, the off-diagonal elements are computed following $a/d^2$ where $d$ is the distance between components on a PIC schematic as shown in Fig.\ \ref{fig:benchmark_photograd} and $a$ is set such that the coefficient relating two phase shifters belonging to the same unit cell is \SI{1}{\%} of the diagonal coefficient. These are realistic crosstalk values. For phase shifters in Mach-Zehnder interferometers, the coefficients describing received crosstalk is either positive or negative depending on the position of the heat emitter. The passive phase vector $\vec{c}_0$ is generated following a Gaussian distribution of mean 0 and standard deviation 0.7 rad. The reflectivity of the beamsplitters is generated according to a Gaussian distribution of mean 0.56 and standard deviation 0.007, closely agreeing with our experimental hardware. Output transmissions are chosen uniformly between 0.7 and 1. The maximum voltage threshold is set to 14 V, allowing all the phase shifters to achieve a $2 \pi$ phase sweep.
The initial values before training of the virtual replica are set to start with a zero vector for the passive phases $\hvec{c}_0$ and a zero matrix for the crosstalk matrix $\hvec{C}_2$. The estimated output transmissions are configured to be 1 and the estimated beamsplitter reflectivity values are set to 0.5.
For each voltage and phase sweep during the characterization, 15 data points are acquired. The learning rates are the following: $\hat{C}_2: 10^{-5}$, $\hvec{R}: 10^{-3}$ and $\hvec{T}_\text{out}: 10^{-3}$. The learning rate scheduling factor is set to 0.7. 

\subsection*{Simulation benchmarking the imperfection mitigation}

For the global optimization compilation, 32 simultaneous phase computations via gradient descent are performed simultaneously following App.\ \ref{app:phase_optim}. Optimization stops for $10^{-6}$ relative variation of both loss function and optimization parameters. For all methods, 32 random relabelings are tested. 

\subsection*{Experimental validation}

We use a single-photon source based on a quantum dot embedded in a cavity, emitting single-photons at \SI{929}{\nano\meter}. A pump laser excites the source with an \SI{80}{\mega\hertz} pulse rate. The 2-photon emission probability is estimated at 1 \%. The stream of emitted single photons is divided into 6 by an active demultiplexer. Fiber delays synchronize the photons such that they enter the 12-mode PIC at the same time. Photons are injected in even input port numbers. An automated mechanical shutter system ensures that only one input port is addressed at a time. The PIC features 126 thermo-optic phase shifters and 132 beamsplitters with fabrication wavelength \SI{940}{\nano\meter}. The PIC reconfiguration time is 2 seconds when reconfiguring all the phase shifters. To speed up the process, we update only voltages that differ by more than \SI{0.1}{\milli\volt} (leading to phase errors below \SI{1e-9}{rad}). We acquire 15 points per phase shifter during voltage and phase sweeps. Photons are detected by 12 superconducting nanowire single-photon detectors and countrates are measured by a correlator. We use a countrate integration time of 1 second. We use the same learning rates and scheduling factor as for the simulation benchmarking to train the virtual replica model. Electric crosstalk is not considered because of an already built-in compensation subroutine by the manufacturer. The characterization protocol carried out the following sequence of stages: V-IFM $\rightarrow$ ML $\rightarrow$ $\phi$-IFM$\rightarrow$ ML $\rightarrow$ ITM. 16500 training and 4125 test samples have been acquired in 17 hours (the data-to-parameter ratio is 1.03). The V-IFM and $\phi$-IFM stages take about 30 minutes. The protocol is run on a i7-12700 \SI{2.1}{\giga\hertz} processor, allowing the ML stages with 500 epochs each to be completed in under 10 minutes. The characterization process took thus about 18 hours.
We measure the unitary amplitude fidelity by compiling 100 Haar-random unitary matrices into phases using the "local correction" method with detector relabeling (see Sec.\ \ref{sec:compilation}). We post-process the acquired matrices by dividing each row by the estimated output transmission. The columns are then normalized such that their elements squared sum to 1.

\section*{Acknowledgements}

The authors would like to express their gratitude to Elianor Hoffmann, Thomas Liege and Francesca Zanichelli for their valuable contributions. We would like to acknowledge Simone Piacentini and Alexia Salavrakos for providing valuable feedback on the manuscript. We sincerely thank Nicolas Heurtel, Alexandre Mercier, Rawad Mezher and Mathias Pont for fruitful discussions. 

This work has been co-funded by the European Commission as part of the EIC accelerator program, under the grant agreement 190188855. We acknowledge funding from the Plan France 2030 through the project ANR-23-PETQ-0013.

\section*{Competing Interests}

A patent has been filed listing A.F., J.S., N.M and N.B. as inventors.

\section*{Author contributions}

Authors are listed in the main title order.
Characterization protocol: A.F., O.F. \& J.S. Protocol generation algorithm: A.F. \& O.F. Imperfection mitigation: A.F. \& J.S. Experimental investigation: A.F., N.M., J.S. \& N.B. Experimental validation: A.F., N.M. \& J.S. Manuscript writing and visualization: A.F. \& N.B. Project supervision: N.M., J.S. \& N.B.

\section*{Data and code availability}

The codebase and data generated as part of this work are available to research groups upon reasonable request from the corresponding authors.

\bibliographystyle{naturemag}
\bibliography{biblio/biblio}

\newpage
\onecolumngrid
\appendix
\section{STATE OF THE ART IN PHOTONIC CHIP CHARACTERIZATION AND OPTIMAL CONTROL}
\label{app:soa}

\begin{table}[]
\begin{tabular}{lclll|ccc|c|}
\cline{6-9}
                                                    & \multicolumn{1}{l}{}                                     &                        &                                                                                                                                                   &                                                                       & \multicolumn{3}{c|}{\begin{tabular}[c]{@{}c@{}}Characterization\\ metrics\end{tabular}}                                                                                                     & \begin{tabular}[c]{@{}c@{}}Control\\ metrics\end{tabular}                \\ \cline{2-9} 
\multicolumn{1}{l|}{}                               & Approach                                                 & \multicolumn{1}{c}{ML} & \begin{tabular}[c]{@{}l@{}}Imperfections taken \\ into account\end{tabular}                                                                       & \multicolumn{1}{c|}{PIC}                                              & \begin{tabular}[c]{@{}c@{}}Circuit\\ $1-\mathcal{F}_a$\end{tabular} & \begin{tabular}[c]{@{}c@{}}Sample \\ cost\end{tabular}       & \begin{tabular}[c]{@{}c@{}}Parameter\\ cost\end{tabular} & \begin{tabular}[c]{@{}c@{}}Unitary \\ $1-\mathcal{F}_a$\end{tabular}       \\ \hline
\multicolumn{1}{|l|}{\cite{Cimini2021},2021}        & \cellcolor[HTML]{000000}{\color[HTML]{FFFFFF} Black-box} & NN                     & all                                                                                                                                               & \begin{tabular}[c]{@{}l@{}}3 modes\\ 2 PS\end{tabular}                & -                                                                 & \begin{tabular}[c]{@{}c@{}}2400\\ (exponential)\end{tabular} & 20750                                                    & -                                                                        \\ \hline
\multicolumn{1}{|l|}{\cite{Youssry2023},2023}       & \cellcolor[HTML]{C0C0C0}Grey-box                         & NN                     & all                                                                                                                                               & \begin{tabular}[c]{@{}l@{}}3 modes\\ 4 PS\end{tabular}                & -                                                                 & 1300                                                         & 450                                                      & 0.26 \%  \footnote{Distribution of unitary matrices is not specified.}  \\ \hline
\multicolumn{1}{|l|}{\cite{Taballione2021},2021}    & Clear-box                                                & No                     & passive phases                                                                                                                                    & \begin{tabular}[c]{@{}l@{}}12-mode CI\\ 126 PS\end{tabular}           & -                                                                 & -                                                            & -                                                        & \begin{tabular}[c]{@{}c@{}}9.6 \% \\ (Haar-random)\end{tabular}          \\ \hline
\multicolumn{1}{|l|}{\cite{Bandyopadhyay2022},2022} & Clear-box                                                & GD                     & \begin{tabular}[c]{@{}l@{}}passive phases, \\ thermal crosstalk,\\ beamsplitter reflectivity, \\ output transmissions\end{tabular}                & \begin{tabular}[c]{@{}l@{}}6-mode CI\\ 36 PS\end{tabular}             & 3.1 \%                                                            & $\geq 23$ \footnote{Voltage sweeps of phase shifters not included.}                                                    & -                                                        & \begin{tabular}[c]{@{}c@{}}1.3 \% \\ (Haar-random)\end{tabular}          \\ \hline
\multicolumn{1}{|l|}{\cite{Taballione2022},2023}    & -                                                        & -                      & -                                                                                                                                                 & \begin{tabular}[c]{@{}l@{}}20-mode CI\\ 380 PS\end{tabular}           & -                                                                 & -                                                            & -                                                        & \begin{tabular}[c]{@{}c@{}}2.6 \% \\ (Haar-random)\end{tabular}          \\ \hline
\multicolumn{1}{|l|}{\textbf{This work}}            & \textbf{Clear-box}                                       & \textbf{GD}            & \textbf{\begin{tabular}[c]{@{}l@{}}passive phases, \\ thermal crosstalk,\\ beamsplitter reflectivity, \\ input/output transmissions\end{tabular}} & \textbf{\begin{tabular}[c]{@{}l@{}}12-mode CI \\ 126 PS\end{tabular}} & \textbf{0.08 \%}                                                  & \textbf{1.4}                                                 & \textbf{1}                                                        & \textbf{\begin{tabular}[c]{@{}c@{}}0.23 \% / 0.15 \% \footnote{Raw value / value with output losses compensation}\\ (Haar-random)\end{tabular}} \\ \hline
\end{tabular}
\caption{
    \textbf{State of the art for photonic integrated circuit (PIC) characterization and optimal control.}
    -: data not available.
    \textbf{Approach)} Black-box: no model for the relation between observed probabilities and applied voltages/electric currents. Gray-box: partial model, where the PIC is not modelled and the measurement obeys the laws of quantum mechanics. Clear-box: full model of the PIC and its imperfections.
    \textbf{ML)} If machine-learning has been used. NN: neural network. GD: model fit by gradient descent.
    \textbf{Imperfections taken into account)} Which PIC imperfections have been considered. Black- and gray-box approaches take by definition all types of imperfections into account, including imperfections that can be hard to model like nonlinear effects on light.
    \textbf{PIC)} Photonic chip used for experimental validation. CI: Clements interferometer, PS: thermo-optic phase shifter.
    \textbf{Circuit $1-\mathcal{F}_a$)} Amplitude infidelity (see App.\ \ref{app:amplitude_fidelity}) between measured output light intensity distributions and corresponding model predictions. Quantifies accuracy of the characterization. The lower, the better.
    \textbf{Sample cost and Parameter cost)} Estimation of sample and computational efficiency. The lower the better. See App.\ \ref{app:soa}.
    \textbf{Unitary $1-\mathcal{F}_a$)} Amplitude infidelity between target unitary matrix and implemented unitary matrix. The lower, the better. Haar-random: target unitary matrices are Haar-randomly chosen. 
}
\label{tab:soa}
\end{table}

We summarize in Table \ref{tab:soa} the state of the art regarding chip characterization and optimal control. We are interested in methods providing the highest standard of characterization and reconfiguration control for PICs, that is by allowing to implement entire unitary matrices by direct dialling. This excludes from our study self-configuration methods limited to light routing from a single input to one output, or partial unitary matrix implementation \cite{Dyakonov2018, Pérez2020, Kondratyev2023}. PIC characterization and control processes should satisfy:

\begin{itemize}
    \item high accuracy of the characterization, here measured by the circuit amplitude fidelity (see App.\ \ref{app:amplitude_fidelity}) which quantifies the difference between the measured output light intensity distributions and the model prediction.
    \item sample efficiency, i.e.\ the model is trained with as few measurements as possible. To assess the sample efficiency of a method, we define 
    \begin{equation}
    \text{sample cost} = \frac{\text{number of training samples}}{n_\text{PS}^2}
    \end{equation}
    where a training sample is an output light distribution measurement used to train the model and $n_\text{PS}$ is the number of phase shifters of the PIC. We have chosen $n_\text{PS}^2$ in the denominator because the dominant imperfection in terms of number of parameters to estimate is thermal crosstalk, which scales as $n_\text{PS}^2$. Hence $n_\text{PS}^2$ is good indicator of the amount of information to gather to characterize a given PIC.
    \item computational efficiency, defined as the processing power required to train a model. We define a metric
    \begin{equation}
    \text{parameter cost} = \frac{\text{number of model parameters}}{n_\text{PS}^2}
    \end{equation}
    which evaluates the number of model parameters to train for a given PIC.
    \item high-fidelity implementation of unitary matrices
\end{itemize}
It is worth noting that when using black- or gray-box approaches for PIC control, it is necessary to train additional machine learning models for imperfection mitigation (e.g.\ the "unitary controller" neural network in \cite{Youssry2023}). 

From Table \ref{tab:soa}, our method yields the highest characterization and unitary implementation accuracy, while guaranteeing scalability thanks to low sample and parameter costs.

The process of characterization inherently requires more computational resources compared to imperfection mitigation. Indeed, imperfection mitigation like references \cite{Hamerly2022, Bandyopadhyay2021} tends to scale linearly with $n_\text{PS}$, whereas PIC characterization with crosstalk depends on $n^2_\text{PS}$. Furthermore, while imperfection mitigation appears to benefit from efficient algorithmic approaches, PIC characterization currently leans on machine learning.

\section{ELECTRIC CROSSTALK}
\label{app:electric_crosstalk}

\begin{figure}[ht!]
    \centering
    \includegraphics{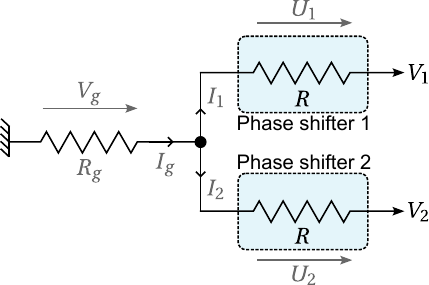}
    \caption{}
    \label{fig:electric_crosstalk}
\end{figure}

We show in this appendix that endowing our virtual replica model with a matrix phase-voltage relation is insufficient for characterizing electric crosstalk and that a third-order-tensor relation is needed for an accurate crosstalk characterisation. We consequently discuss how to tackle the problem from a practical point of view. 
We consider an elementary PIC with two voltage-controlled thermo-optic phase shifters sharing a common ground depicted on Fig. \ref{fig:electric_crosstalk}. The phase shifters can each be modelled by a resistor with resistance $R$, and the ground has a resistance $R_g$. We list our assumptions in this model:
\begin{enumerate}
    \item when taken individually, each phase shifter has a phase-voltage response $\phi(V) \propto V^2/R$ (this is the case for ideal thermo-optic phase shifters, whose response must be proportional to the power dissipated via the Joule effect).
    \item there is no thermal crosstalk between phase shifters
    \item $R$ does not depend on heater temperature
\end{enumerate}

On the considered PIC, an operator applies voltages $V_1$ and $V_2$ on phase shifters 1 and 2 respectively, targeting phase shifts $\phi_1 \propto V_1^2/R$ and $\phi_2 \propto V_2^2/R$ on phase shifters 1 and 2. Because the phase shifters share a common ground, the real voltages $U_1$ and $U_2$ applied on the phase shifters are different from $V_1$ and $V_2$. From Kirchhoff's law and Ohm's law, solving for $U_1$ and $U_2$ yields

\begin{equation}
  \left\{\begin{array}{@{}l@{}}
    U_1 = \frac{(R+R_g)V_1 - R_g V_2}{R+2R_g} \\
    U_2 = \frac{(R+R_g)V_2 - R_g V_1}{R+2R_g}
  \end{array}\right.
\end{equation}
confirming that $V_1 \neq U_1$ and $V_2 \neq U_2$. The phase-voltage relation of each phase shifter is then of the form

\begin{equation}
    \phi = \alpha V_1^2 + \beta V_1V_2 + \gamma V_2^2.
\end{equation}
In the general case of $n_\text{PS}$ on-chip phase shifters, the physical phase applied on phase shifter $i$ is of the form:

\begin{equation}
    \phi^{(i)} = \sum_{k,l} g^{(i)}_{k,l} V^{(k)} V^{(l)}
\end{equation}
where $V^{(k)}$ is the voltage applied on phase shifter $k$. The $n_\text{PS}^3$ numbers $g^{(i)}_{k,l}$ form an order 3 tensor. Trading the matrix phase-voltage relation of the virtual replica for an order 3 tensor relation entails the the number of training data samples for the machine learning stages of the characterization protocol increases from $n_\text{PS}^2$ to $n_\text{PS}^3$. To avoid the $n_\text{PS}^3$ growth of the dataset with increasing PIC size, resource-efficient alternative approaches are to use current-controlled reconfigurable components that are immune to electric crosstalk, or to measure the resistances of the ground resistor and phase shifters resistors to compensate electric crosstalk directly. 

\section{INTERFERENCE FRINGE MEASUREMENTS}
\label{app:ifm}

Interference fringe measurements (IFM) are the standard method for characterizing on-chip phase shifters (PS) (see \cite{Flamini2015} for example). We model the phase-voltage relation of a photonic integrated circuit (PIC) linking the implemented phases $\vec{\phi}$ to the applied voltages $\vec{V}$ as in Sec.\ \ref{sec:modelling} in the main text:
\begin{equation}
    \vec{\phi} = \sum_{k\geq1} C_k \cdot \vec{V}^{\odot k} + \vec{c}_0
\end{equation}
where $\vec{c_0}$ is a vector with $n_\text{PS}$ entries containing the passive phases ($n_\text{PS}$ is the number of on-chip phase shifters), $^{\odot}$ represents element-wise exponentiation, and the $C_k$ are $n_\text{PS} \times n_\text{PS}$ matrices.

In our characterization protocol, we use an initial V-IFM to estimate the passive phases $\vec{c}_0$ and the self-heating coefficients (diagonal elements of the $C_k$ matrices) of each PS. The V-IFM stage consists in sweeping the control voltage of each PS, recording the interference fringe at the circuit output and fitting it with a suitable model. In our characterization process, we also introduce $\phi$-IFMs stages, where each PS is swept in phase from 0 to 2$\pi$ by solving the learned phase-voltage relation. $\phi$-IFMs are performed to update the passive phases stored in the model parameters of the virtual replica.

\subsection{MZIs and meta-MZIs}
\label{app:mzi}

\begin{figure}[!ht]
    \centering
    \includegraphics[width=\textwidth]{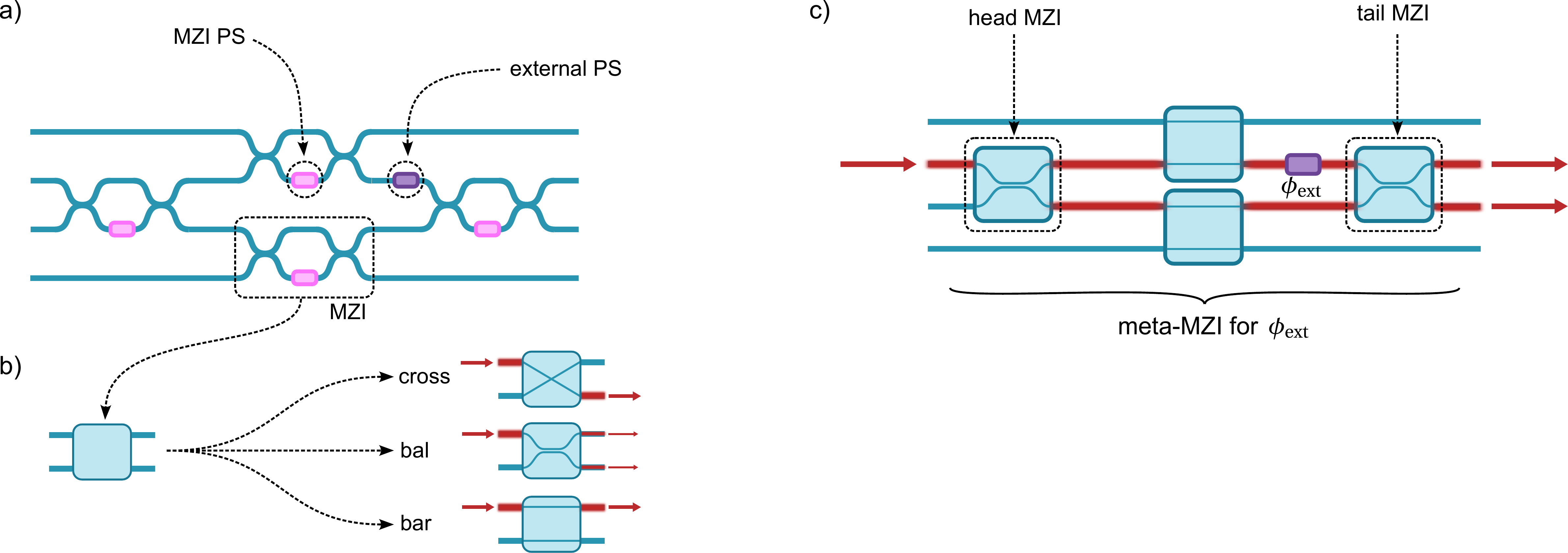}
    \caption{ 
    \textbf{a)} An MZI consists in two beamsplitters and a PS. MZIs are photonic circuit building blocks with 2 inputs and outputs, here represented as blue squares. PS in MZIs are called MZI PS (pink retangles) while others are external PS (purple rectangles).
    \textbf{b)} Depending on the value of the phase, an MZI can realize cross, balanced (bal) and bar configurations. 
    \textbf{c)} We adopt here a schematic view of the same circuit for clarity. A meta-MZI for the external PS $\phi_\text{ext}$ is formed by setting the head and tail MZIs to the balanced configuration. The other MZIs are set to the bar configuration to redirect entirely the light from the head MZI to the tail one. 
    }
    \label{fig:meta-MZI}
\end{figure}

Interferometer meshes feature Mach-Zehnder interferometers, which are photonic circuit building blocks composed of 2 beamsplitters and a PS in-between (see Fig.\ \ref{fig:meta-MZI}a). MZIs act as tunable beamsplitters. Assuming that the MZI beamsplitters have reflectivity 0.5, the MZI can be set to the following configurations depending on the implemented phase on its MZI PS (see Fig.\ \ref{fig:meta-MZI}b):

\begin{itemize}
    \item \emph{cross} configuration: for $\phi=0$, the MZI performs a perfect swap between the two input modes
    \item \emph{balanced} (bal) configuration: for $\phi=\pi/2$, the MZI acts like a symmetric beamsplitter
    \item \emph{bar} configuration: for $\phi=\pi$, the MZI acts like a perfect mirror    
\end{itemize}

On-chip reconfigurable PSs are either categorized as MZI PSs, enclosed in an Mach-Zehnder, or external PSs (see Fig.\ \ref{fig:meta-MZI}a).  Still, an external PS can be used to form an MZI by setting a preceding MZI (named head MZI) and a succeeding MZI (named tail MZI) to the balanced configuration (see Fig.\ \ref{fig:meta-MZI}c), hence mimicking the structure of a standard MZI. Such a circuit setting is called meta-MZI. If other MZIs are present on the light path between the head and tail MZI, their configuration must be chosen such that the light is routed from the head MZI to the tail MZI.

\begin{figure}[h]
    \centering
    \includegraphics[width=\textwidth]{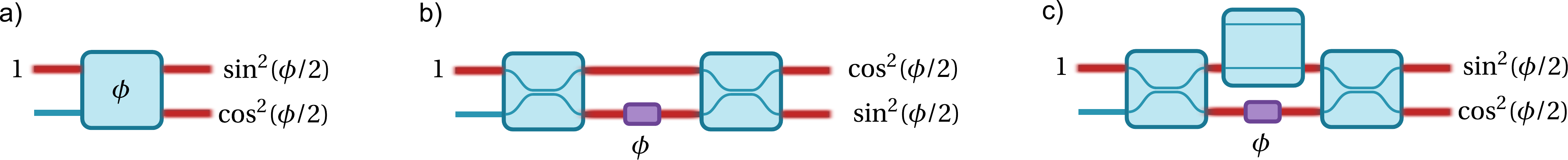}
    \caption{Light splitting ratio as a function of the implemented phase for an MZI and different meta-MZIs}
    \label{fig:meta_mzi_splitting}
\end{figure}

We now discuss the dependence of the light splitting ratio as a function of the implemented phase for MZIs and meta-MZIs. The unitary matrix associated to an MZI from Fig.\ \ref{fig:meta-MZI} is

\begin{equation}
    i e^{\frac{i\phi}{2}}
    \begin{bmatrix}
    -\sin(\frac{\phi}{2}) & \cos(\frac{\phi}{2}) \\
    \cos(\frac{\phi}{2}) & \sin(\frac{\phi}{2})
    \end{bmatrix}.
\end{equation}
Hence, when light enters from the MZI top mode, a fraction $\cos^2(\phi/2)$ is transmitted to the bottom mode (see Fig.\ \ref{fig:meta_mzi_splitting}). For a meta-MZI consisting in 2 MZIs in the balanced configuration, the transmitted fraction of light is $\sin^2(\phi/2)$. Inserting an additional MZI in the bar configuration on the top mode reverts the transmitted fraction back to $\cos^2(\phi/2)$. Consequently, meta-MZIs exhibit an oscillating output light intensity when sweeping their associated external phase that is reminiscent of standard MZIs. The exact dependence of the splitting ratio of a meta-MZI is however determined by the number of traversed components between the head and tail MZIs.

\subsection{Characterization of phase shifters with interference fringe measurements}
\label{app:ifm_principle}

We present here the different steps for characterizing a given phase shifter denoted PS$_i$.

\begin{figure}[h]
    \centering
    \includegraphics[scale=0.3]{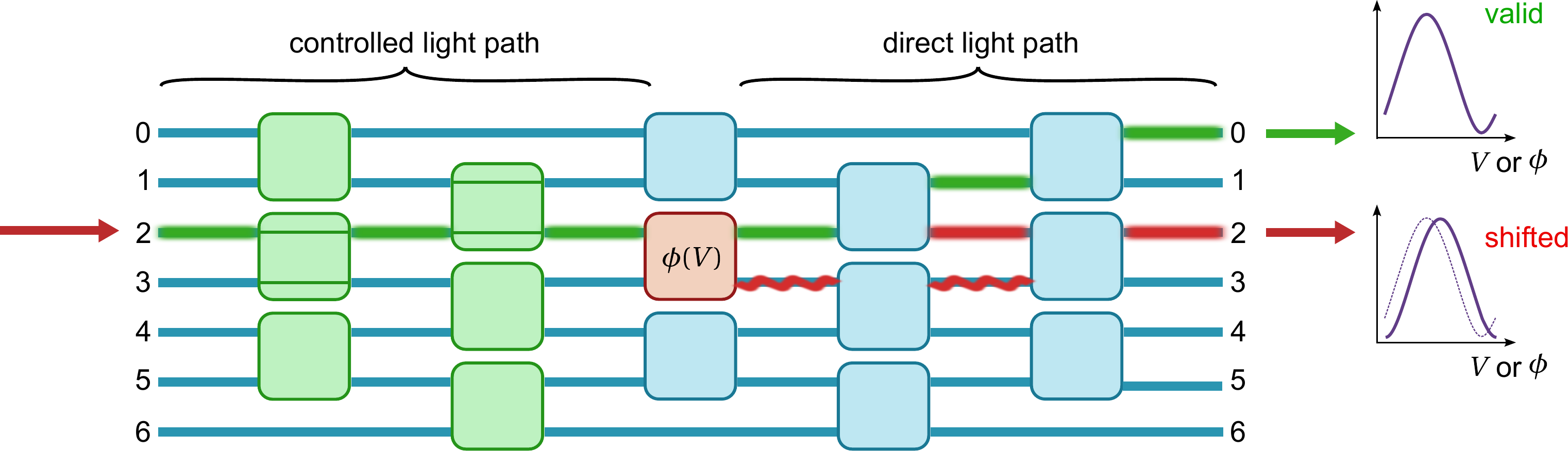}
    \caption{
    \textbf{Guidelines for light routing inside the PIC.}
    Squares represent MZIs. The PS of green MZIs has been characterized, hence green MZIs can be controlled and set to the bar configuration. The red MZI contains the PS that is to be characterized, with an unknown phase-voltage relation $\phi(V)$. The green light path represents a valid light route to characterize $\phi(V)$. Light is injected in input 2, two characterized MZIs are set to the bar configuration to route all the light to the red MZI. Starting from the red MZI, we do not control light routing as the blue MZIs are unknown. We monitor output 0 because the light path between the red MZI and the output is direct, i.e.\ uncontrolled light emitted by the red MZI is not able to recombine with itself on the green path. On the contrary, the red light path is not direct. Red wavy light interferes with the red light path, leading to a shifted measured interference fringe on output 2. 
    }
    \label{fig:routing_example}
\end{figure}

\begin{enumerate}
    \item \textbf{Light routing.} Given an input for injecting light into the photonic circuit and an output port monitored by a detector, we set up a route in the circuit such that the light passes through the MZI or chosen meta-MZI containing PS$_i$. To set up this route, bar or cross configurations are implemented on previously characterized MZIs. Appropriate routes must not allow the injected light to recombine with itself in an uncontrolled way. Parasitic recombinations cause the observed interference fringe to shift, hence leading to imprecise measurements. We call "direct paths" light routes on which it is not possible for light to interfere with itself. Alternatively, light can be safely routed through already characterized MZIs. We show on Fig.\ \ref{fig:routing_example} a light routing example. 
    \item \textbf{Voltage/phase sweep} Once the route has been established, the voltage or phase of PS$_i$ is swept and the measured light intensities are saved for processing. In the case of a phase sweep, the voltages to apply on the PIC are computed from the currently known phase-voltage relation stored in the virtual replica model parameters. Only PS$_i$ and the subset of active PSs used to route light are taken into account when solving the phase-voltage relation, in order to keep crosstalk levels and the number of heating PSs low. 
    \item \textbf{Measured data processing} 
    We assume for clarity a phase-relation $\vec{\phi} = C_2 \cdot \vec{V}^{\odot2} + \vec{c}_0$.
    \begin{itemize}
        \item 
        For a \textbf{voltage} sweep, the measured oscillation is of the form
        \begin{equation}
            f(V) = a_i \cos^2\left(\frac{C_2^{(i,i)} V^2 + c_0^{(i)} - \theta_i}{2}\right) + b_i
        \label{eq:fit_function}
        \end{equation}
        where $C_2^{(i,i)}$ is the self-heating coefficient, $c_0^{(i)}$ is the passive phase, $\theta_i$ is an optional phase offset, and $a_i$ and $b_i$ are additional fit parameters. The phase offset $\theta_i$ is computed in advance taking into account the light routing input and output port of the MZI or meta-MZI containing PS$_i$. If PS$_i$ is an external PS, the phase offset $\theta_i$ also depends on the structure of associated meta-MZI (see App. \ref{app:mzi}).

        Fitting the data yields an estimation of the passive phase $c_0^{(i)}$ and self-heating coefficient $C_2^{(i,i)}$.
        \item For a \textbf{phase} sweep, we use two methods for processing the data. The \textbf{fast} method generates the expected output $f(\phi)$ using the beamsplitter reflectivities and output transmissions currently learned by the virtual replica. The data points are fitted with $a+b \cdot f(\phi+\delta\phi)$, where the fit parameters are $a, b$ and $\delta\phi$. The passive phase of PS$_i$ is then updated from $c_0^{(i)}$ to  $c_0^{(i)} + \delta\phi$. The fit quality of the fast method, that is the distance between the data points and the fit curve, will stagnate starting from a certain (ML + $\phi$-IFM) characterization iteration (see Sec.\ \ref{sec:p_ifm}). When the fit quality starts to stagnate with the fast method, the \textbf{precise} fit protocol is used. For the precise fit, an optimization is run to find the passive phase update $c_0 + \delta\phi$ such that the generated output curve reproduces best the measured data.
    \end{itemize}
\end{enumerate}

\subsection{Automated characterization protocol}
\label{app:charac_algo}

Establishing the IFM protocol, that is the sequence of characterization for the on-chip PS, as well as the associated light routes, input ports and output ports, is straightforward enough to be implemented by hand on small PICs, but as the PICs grow in size and complexity, it becomes impractical. We designed an algorithm that generates the IFM protocol from the PIC logical layout and the used input ports. We describe in the following how the algorithm works.

\begin{figure}[h]
    \centering
    \includegraphics[width=\textwidth]{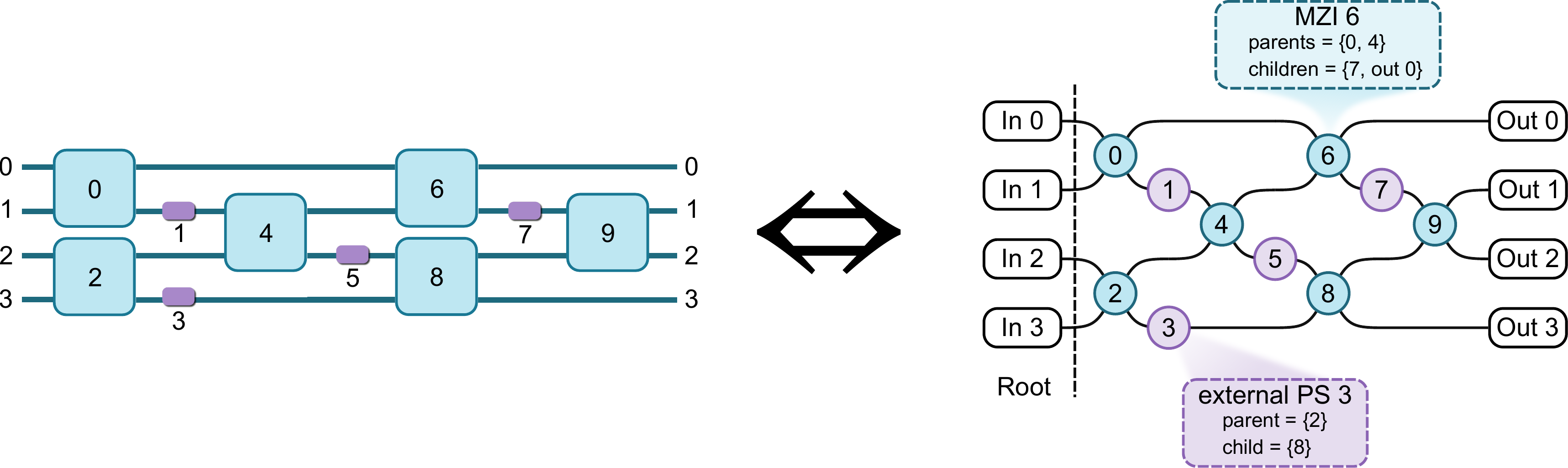}
    \caption{
    The PIC on the left is equivalent to its graph representation $\mathcal{G}$ on the right. The nodes of the graph can be MZIs, external PS or input/output ports.
    }
    \label{fig:pic_graph}
\end{figure}

We denote $m$ the number of input and output ports of the PIC to characterize. First, the PIC is translated into a graph $\mathcal{G}$ with $m$ root input nodes as in Fig.\ \ref{fig:pic_graph}. Each node in the graph is either an MZI, an external PS or an input/output port. The PIC layout dictates the arrangement of the graph nodes and edges. The direction of light propagation in the PIC defines upstream and downstream components. Each node has as following attributes:

\begin{itemize}
    \item \texttt{parents}, the indices of the nodes directly upstream of the node
    \item \texttt{children}, the indices of the nodes directly downstream of the node
\end{itemize}
To establish the IFM protocol of the PIC, the algorithm generates first the protocol for the MZI PSs, then appends the sequence for the external PSs. The same IFM protocol is used for V-IFMs and $\phi$-IFMs. 

\subsubsection{MZI characterization}

\begin{figure}[h]
    \centering
    \includegraphics[width=\textwidth]{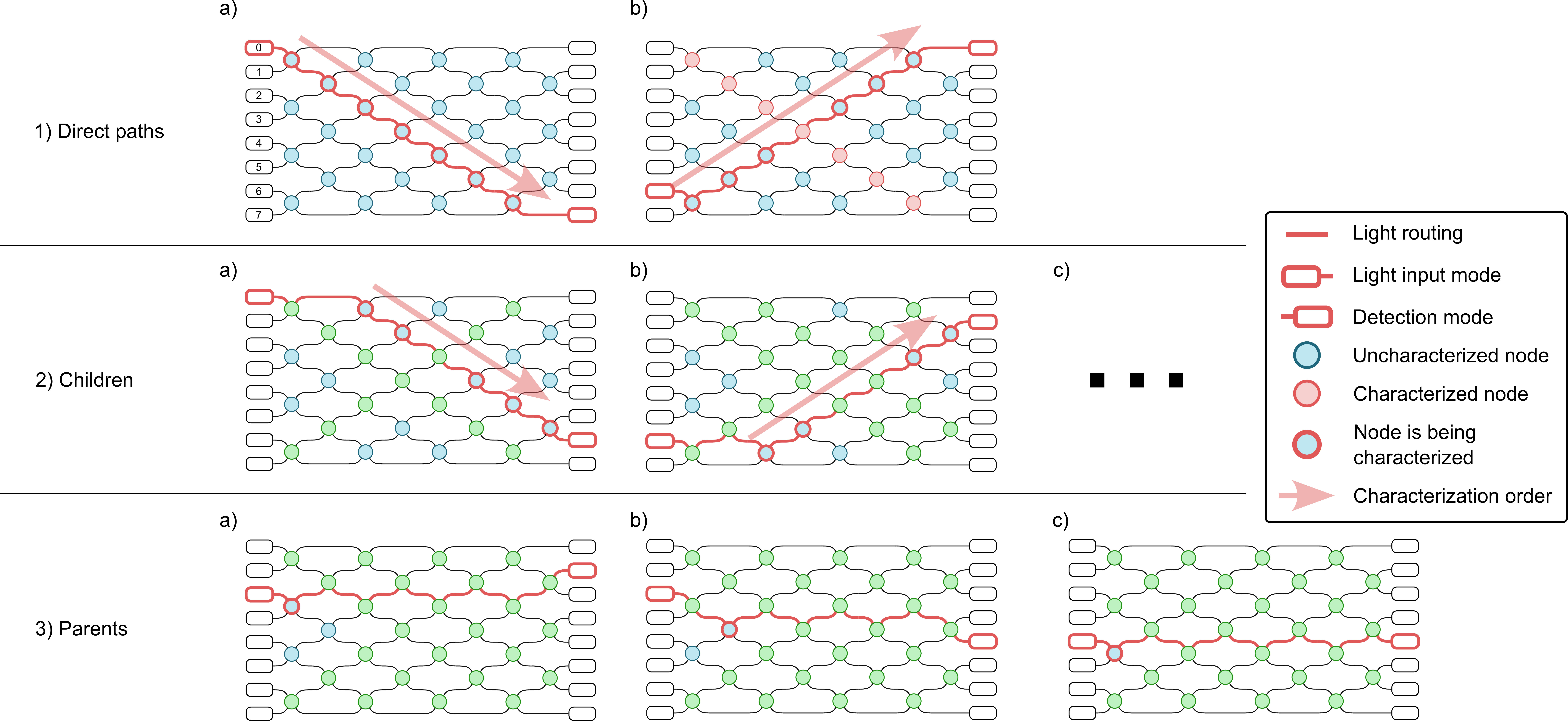}
    \caption{
    \textbf{MZI characterization procedure for an 8-mode Clements interferometer.} 
    External PSs are not shown for clarity. Blue notes indicate nodes that have not been characterized yet. Green nodes have been characterized. Light is only injected in even input ports. \textbf{1)} The algorithm starts by finding the direct light paths of the PIC and characterizes the nodes on the path. Thick red lines indicate the light input port, light routing, nodes being characterized, and detection output port. The red transparent arrow gives the order of characterization. \textbf{2)} Subsequently, the children of the characterized nodes are processed. We only show two steps out of 6 for brevity. \textbf{3)} Finally, the parents of the direct path nodes are characterized. 
    }
    \label{fig:mzi_charac_example}
\end{figure}

\begin{figure}[h]
    \centering
    \includegraphics[width=0.4\textwidth]{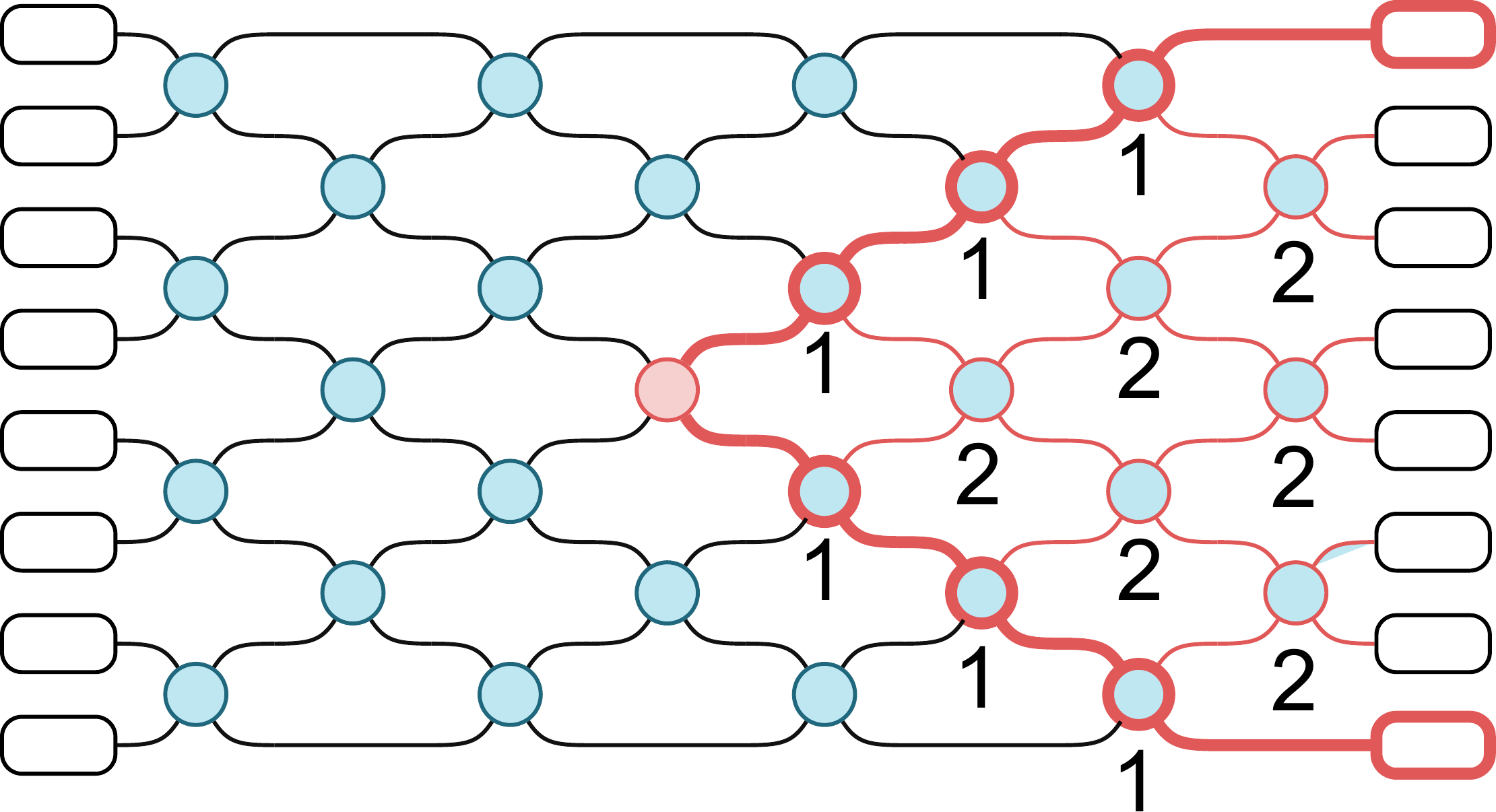}
    \caption{
    \textbf{Method for finding direct light paths.} 
    Blue circles are MZI nodes. To find a direct light path that extends from the middle red node to the PIC outputs, a beam of light is propagated starting from that MZI node. At each encountered MZI node, the beam splits in 2. The number under each node is the \texttt{illumination} attribute, which counts the number of incident beams on the node. Once the beam reaches the output, the algorithm backtracks the paths formed by nodes with illumination 1. The suitable paths are marked by a thick red line. Light travelling along these paths is not affected by parasitic interferences with other parts of the beam.
    }
    \label{fig:direct_light_path}
\end{figure}

The procedure for MZI characterization is shown on Fig.\ \ref{fig:mzi_charac_example} in the case of an 8-mode Clements interferometer. 
\begin{enumerate}
    \item \textbf{Direct path} The algorithm starts by finding the direct light paths of the PIC, and characterizes the PSs corresponding to the MZI nodes along these paths. This is a reliable method for characterizing PSs within an uncharacterized circuit. Direct light paths guarantee that the injected light does not interfere with itself in an uncontrolled manner, which would offset the measured output fringe and lead to wrong passive phase estimation. 
    \item \textbf{Children} The nodes characterized in step 1 can now be set to bar or cross configuration. Hence, light can be routed to their child nodes. The algorithm determines the output port to monitor by finding direct paths starting from the child nodes. This procedure is iterated to characterize the next generation of child nodes.
    \item \textbf{Parents} Once there are no descendants left in step 2, the algorithm characterizes the parent nodes of the direct path nodes characterized in step 1. The procedure is equivalent to the step 2. The light routes are established by letting the algorithm propagate light backwards from the output to the input ports. The procedure is iterated until all nodes are characterized.
\end{enumerate}

When routing light through characterized nodes, we use bar configurations. In our experimental regime where the 928 nm light operation wavelength is strongly detuned from the 940 nm operation wavelength, the bar configuration is more faithfully implemented  by MZIs than the cross configuration. In the case where random fabrication errors dominate, cross configurations should be preferred. 

The protocol generation algorithm requires a function that determines direct light paths in the PIC. To find such paths, each node in the graph $\mathcal{G}$ is provided with an integer attribute called \texttt{illumination} counting the number of incident light beams on the node. The illumination of each node is computed as displayed on Fig.\ \ref{fig:direct_light_path}. By backtracking the nodes \texttt{illumination} 1, the function establishes the direct light paths starting from a given node.

We have also implemented a function that finds light routes through characterized MZI nodes. To determine the light routes for parent nodes (step 3 on Fig.\ \ref{fig:mzi_charac_example}), we use the same functions, but with light propagating backwards in the circuit.

\subsubsection{External PS characterization}

\begin{figure}[h]
    \centering
    \includegraphics[width=0.8\textwidth]{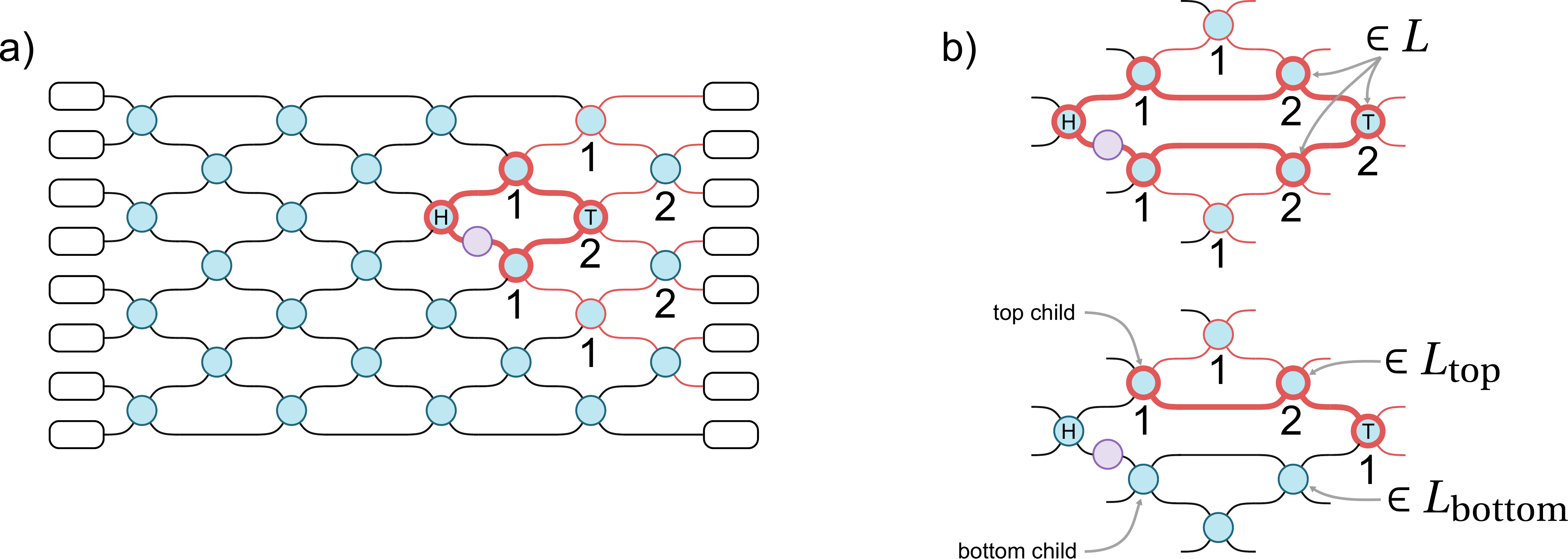}
    \caption{
    \textbf{Meta-MZI determination procedure}
    Blue circles are MZIs. Purple circle is the external PS to characterize. Blue circles with thin red outline are illuminated MZI nodes by the search beam. Blue circles with thick red outline are MZI nodes illuminated by the search beam that participate in the adequate meta-MZI.  Thick red waveguides participate in the meta-MZI while thin red waveguides are simply illuminated.
    \textbf{a)} In the case of a Clements interferometer, the head MZI (H) is simply the parent of the external PS and the tail-MZI is the first MZI with \texttt{illumination} 2 (T).
    \textbf{b)} We illustrate our method in a more complex mesh. The appropriate head MZI is marked (H) and the tail MZI is marked (T) Top: Light is propagated from the first parent of the external PS. The MZIs with \texttt{illumination} 2 are stored in a list $L$.
    Bottom: Light is propagated from the top child of the head MZI. The MZI with \texttt{illumination} 2 is stored in $L_\text{top}$. We do the same with the bottom child and $L_\text{bottom}$. The searched tail MZI is the element from $L$ that does not belong to $L_\text{top}$ and $L_\text{bottom}$ and that is closest to the head-MZI.
    }
    \label{fig:meta_mzi_find}
\end{figure}

Once all the MZIs have been characterized, we turn our attention to the remaining external phase shifters. To characterize an external PS, it must be enclosed in an appropriate meta-MZI (see App.\ \ref{app:mzi}).

Our algorithm constructs appropriate meta-MZIs for a given external PS and uses the shortest one to perform the characterization, that is the one involving the fewest MZI nodes. The method is illustrated in Fig.\ \ref{fig:meta_mzi_find}.  The algorithm tests if the closest MZI node upstream of the external PS can be used as a head MZI for a meta-MZI by propagating a beam of light from that MZI node and retrieving the list $L$ of MZI nodes with \texttt{illumination} attribute 2. The appropriate tail MZI is the element of $L$ that collects two direct paths of light emanating from the head MZI, with only one of them traversing the external PS. $L$ may contain MZI nodes that recombine light emanating from only one of the outputs of the head MZI (see Fig.\ \ref{fig:meta_mzi_find}b). To get rid of them, we collect the list $L_1$ (resp. $L_2$) of MZIs with \texttt{illumination} 2 obtained by propagating light from the top (resp. bottom) child of the head MZI. The list of suitable tail MZIs for the meta-MZI is then $L\setminus\{L_1 \cup L_2$\}. This method works with more general meshes than Clements interferometers. If the first parent of the external PS to characterize does not yield any suitable meta-MZI, the algorithm tries further ancestors.

If the meta-MZI includes other external PSs, they must have been already characterized and held at zero radians while the PS of interest is swept. This gives the order in which the external PSs are characterized.

\subsection{Impact of imperfections on measurement accuracy}
\label{app:ifm_imperfect}

We examine the impact of PIC imperfections on estimations of the passive phase $\vec{c}_0$ in a V-IFM. Differential output transmissions do not affect the measurement, because the fringe is measured in raw detector units. This is different from the $\phi$-IFM case, where the output intensity distribution is normalized to sum to 1. Thus we will consider only hardware beamsplitter reflectivity errors and thermal crosstalk.

\subsubsection{Beamsplitter reflectivity errors}

\begin{figure}[h]
    \centering
    \includegraphics[width=\textwidth]{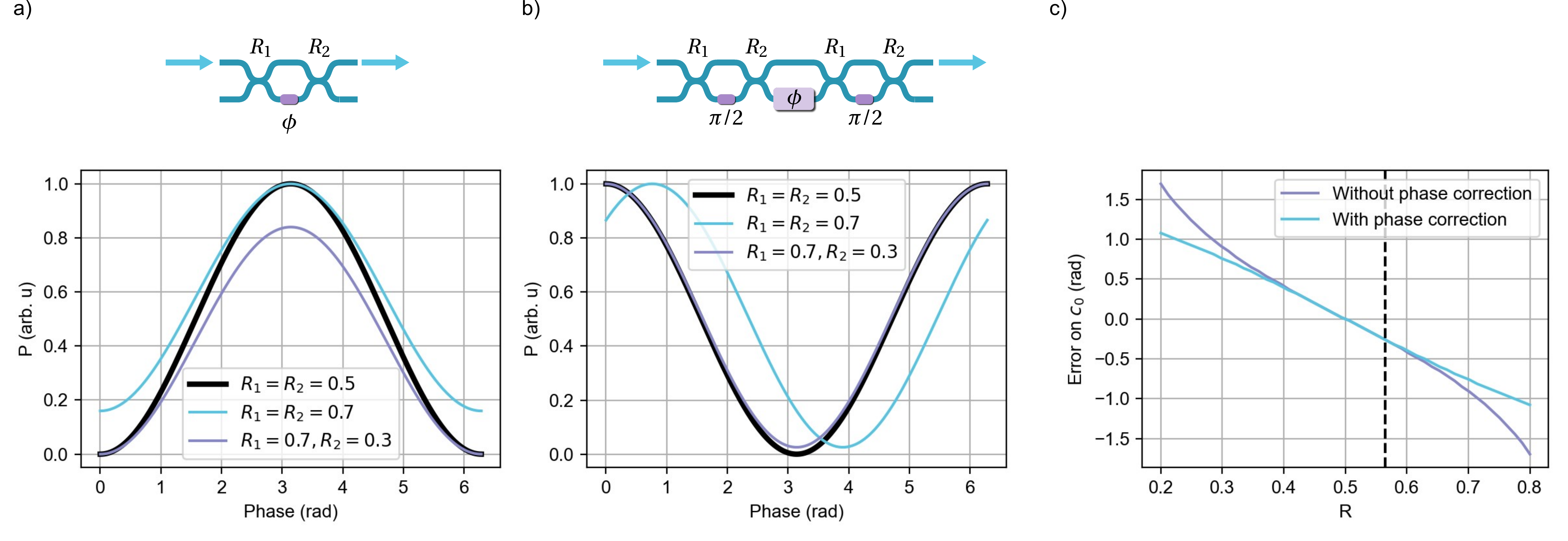}
    \caption{
    \textbf{Impact of beamsplitter reflectivity errors on passive phase measurement}
    \textbf{a)} We simulate the interference fringe of an MZI whose beamsplitters have reflectivities $R_1$ and $R_2$, the ideal case being $R_1=R_2=0.5$. We input and detect on the top mode. If $R_1=R_2 \neq 0.5$ (systematic error) or $R_1=1-R_2$ (symmetric fabrication error around 0.5), the fringe is not displaced and the passive phase is correctly measured.
    \textbf{b)} Here we simulate the same process but for an external PS between two MZIs, forming a meta-MZI as is required to characterize the external PS. For errors of the type $R_1=1-R_2$ the fringe does not shift, contrary to $R_1=R_2 \neq 0.5$. This entails errors in the estimation of the passive phase.
    \textbf{c)} We plot this measurement error on the passive phase in the setting $R_1=R_2=R \neq 0.5$ as a function of $R$ for two cases. In the first case, we apply a phase $\pi/2$ on the two MZIs enclosing the external PS. In the second case, we correct the MZI phases such that they continue acting like symmetric beamsplitters even if $R \neq 0.5$. This procedure reduces measurement errors significantly only when $R$ deviates strongly from 0.5. The dashed line represents the average reflectivity value on the PIC we used to validate experimentally our process (Sec.\ \ref{sec:ascella} in main text), causing a systematic error of -250 mrad on the measured passive phase.
r    }
    \label{fig:refl_impact}
\end{figure}

Beamsplitter reflectivity errors do not affect the measurement of the passive phase of MZI PSs as simulated in Fig.\ \ref{fig:refl_impact}a, but they do affect the measurement for an external PS which is enclosed in a meta-MZI (see App.\ \ref{app:mzi}). The interference fringe is displaced by systematic beamsplitter reflectivity errors (see Fig.\ \ref{fig:refl_impact}b), leading to a wrong estimation of the passive phase in practice for external PS. The value of this error is plotted as a function of reflectivity on Fig.\ \ref{fig:refl_impact}c.

\subsubsection{Thermal crosstalk}

When performing a voltage sweep on a designated PS, light is routed from a PIC input to the PS and then to an output port. Hence some components emit heat while the voltage is swept to achieve this routing. This causes the fringe to be displaced, because the observed phase $\tilde{\phi}_j$ for PS $j$ is

\begin{equation}
    \tilde{\phi}_j = \phi_j + \sum_{i \neq j} (C_2)_{j,i} V^2_i
\end{equation}
where the first term $\phi_j$ is the one we want to observe and the second term is a parasitic shift due thermal crosstalk. We estimate the impact by running an IFM simulation with the same parameters as our simulations described in Methods on a 12-mode Clements interferometer. The average estimation error on the passive phase is 100 mrad, following the protocol generated by our algorithm described in App.\ \ref{app:charac_algo} which already seeks to minimize the number of active components.

\section{PHOTONIC CHIP ACCURACY METRICS}
\label{app:pic_metrics}

We expose in this section several metrics used to assess the amount of control over a photonic integrated circuit (PIC).

\subsection{Total variation distance}
\label{app:tvd}

We use the total variation distance (TVD) to quantify the difference between two PIC output light intensity distributions $\vec{p}$ and $\vec{q}$, which are vectors whose components sum to one. The TVD is computed as
\begin{equation}
    \text{TVD}(\vec{p}, \vec{q}) = \frac{1}{2} \sum_i |p_i - q_i|
\end{equation}
The TVD expresses the amount of difference between $\vec{p}$ and $\vec{q}$ with a number between 0 ($\vec{p}=\vec{q}$) and 1 (maximal difference between $\vec{p}$ and $\vec{q}$). The TVD is more natural than the mean square error (MSE) for apprehending the amount of difference between an output light intensity distribution acquired on a physical PIC and the corresponding virtual replica prediction. We use MSE nevertheless as the cost function for the gradient descent in the ML characterization stage (see Sec.\ \ref{sec:ml}). In the main text, to assess the quality of the virtual replica training, we monitor the average TVD achieved on the test dataset, denoted $\text{TVD}_\text{test}$.

\subsection{Fidelity}
\label{app:fidelity}

It is essential in our study of PIC accuracy to be able to assess the difference between a unitary matrix $U$ implemented by a PIC with $m$ modes and a target unitary matrix $V$. A straightforward method for accomplishing this task involves employing a matrix inner product, and in this context, we adopt the Frobenius inner product \cite{Hiai2014}
\begin{equation}
    \langle U , V \rangle _ F = \text{Tr}\left(U^\dagger V\right)
\end{equation}
The Frobenius inner product is complex-valued, hence we fabricate a real quantity which we call the fidelity
\begin{equation}
    \mathcal{F}\left(U, V\right) =   \frac{\left| \langle U , V \rangle _ F \right|}{m} 
\end{equation}
The infidelity is defined as $1-\mathcal{F}$. We show that the fidelity is a number between 0 and 1 and we provide the condition that leads to achieving the maximum fidelity of 1.
\\
\hrule
\begin{prop} Let $U$ and $V$ be $m\times m$ unitary matrices.
\begin{itemize}
    \item $ 0 \leq \mathcal{F}\left(U, V\right) \leq 1$
    \item $ \mathcal{F}\left(U, V\right) = 1 \Leftrightarrow U = e^{i\theta}V$
\end{itemize}
\end{prop}
\hrule
\begin{proof}
    We use the Cauchy-Schwarz inequality for the Frobenius inner product yielding
    \begin{equation}
        \left|  \langle U , V \rangle _ F \right| \leq \| U  \|_F \cdot \| V \|_F
    \end{equation}
    where $\|\cdot\|_F$ is the norm induced by the Frobenius inner product. We compute
    \begin{equation}
        \| U \|_F 
        = \sqrt{\langle U , U \rangle _ F} 
        = \sqrt{\text{Tr}\left(U^\dagger U\right)} 
        = \sqrt{\text{Tr}\left( I_m\right)}
        = \sqrt{m}
    \end{equation}
    where we have used the unitarity of $U$ for the third equality and $I_m$ is the identity matrix. The same applies for $V$. Hence,
    \begin{equation}
        \left|  \langle U , V \rangle _ F \right| \leq m
    \end{equation}
    We have proven the first assertion. The Cauchy-Schwarz inequality is saturated if and only if
    \begin{equation}
        U = \lambda V
    \end{equation}
    with $\lambda$ a complex scaling factor. Because $U$ and $F$ have the same Frobenius norm, this implies $|\lambda|=1$. This proves the second point.
\end{proof}
\hrule

\subsection{Amplitude fidelity}
\label{app:amplitude_fidelity}

Measuring the amplitude of the elements of the matrix $U$ implemented by a PIC is straightforward. For each PIC input:
\begin{enumerate}
    \item Measure the output light intensity distribution $\vec{p}$
    \item Normalize $\vec{p}$ such that its elements sum to 1
    \item Compute $\sqrt{\vec{p}}$, where the square root is applied element-wise
\end{enumerate}
The resulting vector yields the amplitudes of the elements of the $i^\text{th}$ column of $U$. In contrast, accessing the phase of these elements is a more complex, time-consuming and error-prone experimental process. Consequently, it is widely preferred in hardware experiments to measure and work with the matrix $|U|$, where the absolute value operator is applied element-wise. In particular, the elements of $|U|$ are non-negative and the sum of the squares of each column of $|U|$ sums to 1. Consequently, the metric of choice in the litterature denoted "amplitude fidelity" for evaluating the difference between two matrices $P$ and $Q$ satisfying these two properties is given by
\begin{equation}
    \mathcal{F}_a\left(P, Q\right) = \frac{\text{Tr}\left(P^T Q\right)}{m}
\end{equation}
The amplitude infidelity is defined as $1-\mathcal{F}_a$. We show in the following how the amplitude fidelity compares to the fidelity. We also prove that the amplitude fidelity between two matrices is maximal if and only if these matrices are equal.
\\
\hrule

\begin{prop}
    $\mathcal{F}\left(U, V\right) \leq \mathcal{F}_a\left(|U|, |V|\right)$
\end{prop}

\hrule

\begin{proof}
    The inequality is readily obtained from
    \begin{equation}
        \mathcal{F}\left(U, V\right) 
        =\frac{1}{m} \left|\sum_{i,j} u_{i,j}^* v_{i,j}\right|
        \leq
        \frac{1}{m} \sum_{i,j} |u_{i,j}| |v_{i,j}|
        = \mathcal{F}_a\left(|U|, |V|\right)
    \end{equation}
\end{proof}

\hrule

\begin{prop}
    Let $P$ and $Q$ be $m\times m$ matrices whose elements are non-negative and whose columns sum to 1 when squared element-wise.
    \begin{itemize}
        \item $ 0 \leq \mathcal{F}_a\left(P, Q\right) \leq 1$
        \item $\mathcal{F}_a\left(P, Q\right)=1 \Leftrightarrow P=Q$
    \end{itemize}
\end{prop}

\hrule

\begin{proof}
    The proof follows a similar path as the proof for the fidelity. The amplitude fidelity is related to the Frobenius inner product by
    \begin{equation}
    \mathcal{F}_a\left(P, Q\right) =   \frac{\langle P , Q \rangle _ F }{m} 
    \end{equation}
    We use the Cauchy-Schwarz inequality for the Frobenius inner product yielding
    \begin{equation}
        \langle P , Q \rangle _ F = \left|  \langle P , Q \rangle _ F \right| \leq \| P  \|_F \cdot \| Q \|_F
    \end{equation}
    where $\|\cdot\|_F$ is the norm induced by the Frobenius inner product. We compute
    \begin{equation}
        \| P \|_F 
        = \sqrt{\langle P , P \rangle _ F} 
        = \sqrt{\text{Tr}\left(P^T P\right)} 
        = \sqrt{\sum_{j=1}^m \left(\sum_{i=1}^m p_{i,j}^2 \right)}
        = \sqrt{\sum_{j=1}^m 1}
        = \sqrt{m}
    \end{equation}
    where we have used the fact that the elements squared belonging to a same column of $P$ sum to 1 by assumption. The same applies for $Q$. Hence,
    \begin{equation}
        \left|  \langle P , Q \rangle _ F \right| \leq m
    \end{equation}
    We have proven the first assertion. The Cauchy-Schwarz inequality is saturated if and only if
    \begin{equation}
        P = \lambda Q
    \end{equation}
    with $\lambda$ a real scaling factor. Because $P$ and $Q$ have the same Frobenius norm, this implies $\lambda=1$. This proves the second point.
\end{proof}
\hrule
\vspace{0.3cm}

Restricting to amplitudes is also beneficial, as the matrix implemented by a PIC is not unitary in general due to differential optical input and output transmissions. We only demand from acquired amplitude matrices to have normalized columns. There is however a caveat when restricting to the amplitudes. Measuring an amplitude fidelity $\mathcal{F}_a\left(|U|, |V|\right)=1$ does not automatically imply $U=V$ in the general case. But in the context of matrices produced by PICs, the arguments of the elements of the matrices $U$ and $V$ are directly related to their amplitude. Hence, one can reasonably assume that achieving an amplitude fidelity of 1 with a PIC should correlate to a very good fidelity, if not also equal to 1. 

Experimentally, we do not measure the implemented matrix $|U|$ on the PIC in its entirety. This is because our setup sends photons in every other input. Hence, the denominator $m$ of $\mathcal{F}_a$ is replaced with the number of addressed inputs, which is 6 in our case.

\subsection{Other metrics in the literature}

Reference \cite{Bandyopadhyay2021} quantifies the difference between two $m \times m$  unitary matrices $U$ and $V$ using the Frobenius norm
\begin{equation}
    \varepsilon = \frac{\| U - V \|_F}{\sqrt{m}}.
\end{equation}
We can relate it to the Frobenius inner product via
\begin{equation}
    \text{Re} \left\{ \frac{\langle U, V \rangle_F}{m}    \right\} = 1 - \frac{\varepsilon^2}{2}.
\end{equation}
In order of magnitude, we can relate it to our definition of fidelity
\begin{equation}
    \mathcal{F} \approx 1 - \frac{\varepsilon^2}{2}.
\end{equation}
Reference \cite{Youssry2023} works with the fidelity defined as $(\mathcal{F}_a)^2$, and normalizes the acquired PIC amplitude matrices such that rows and columns sum to 1 when their elements are squared.

\section{SIMULATED CHARACTERIZATION OF A 24-MODE CLEMENTS INTERFEROMETER}
\label{app:max_modes}

\begin{figure*}
    \centering
    \includegraphics[width=\textwidth]{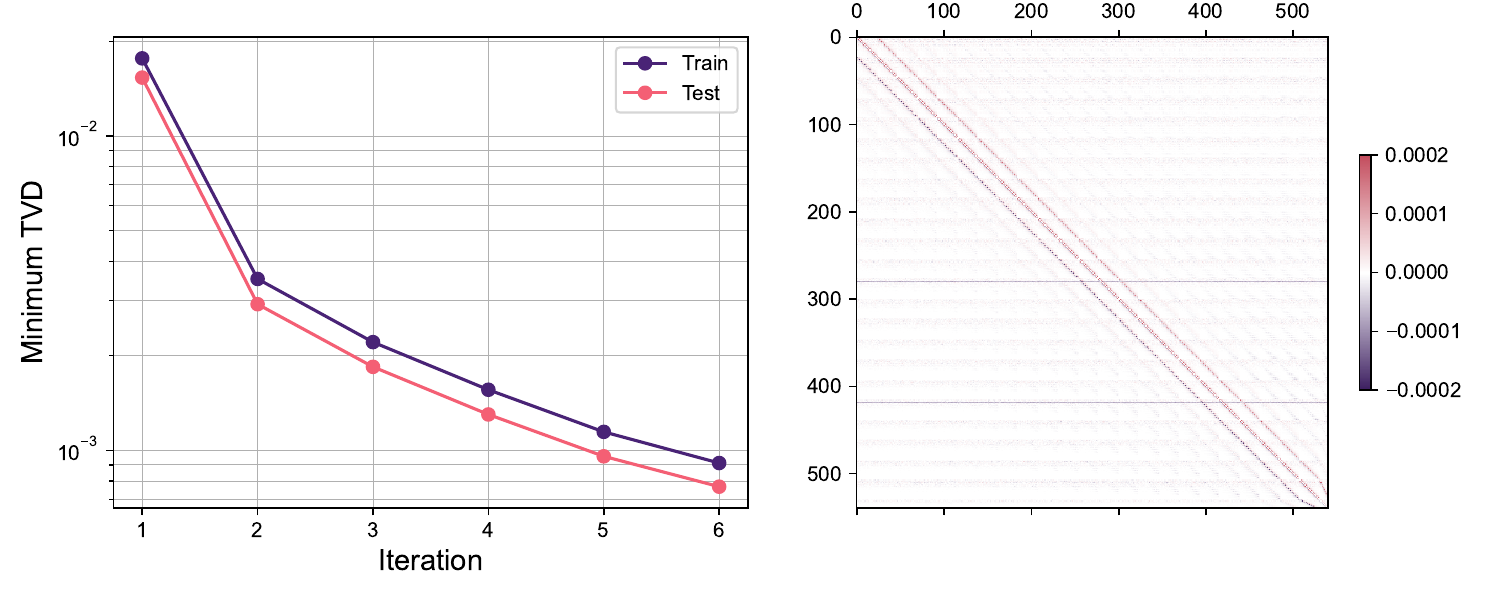}
    \caption{
        \textbf{Characterization of a simulated 24$\times$24 Clements interferometer mesh using our protocol. Left)} Minimum of reached TVD evaluated on test and training dataset for each (ML-IFM) iteration. \textbf{Right)} Estimated crosstalk matrix $\hat{C}_2$. The values have been clipped to [-0.002, 0.002] for clearer visualization.
    }
    \label{fig:max_modes}
\end{figure*}

We carried out the simulation of a Clements interferometers on 24-modes, with crosstalk, beamsplitter reflectivity errors and inhomogeneous output losses. The parameters of the simulated physical device are initialized following the corresponding section in Methods, except the beamsplitter reflectivity values are \SI{0.52(1)}{} and the crosstalk strength is 0.5 \%. The learning rates have been modified to: $\hat{C}_2: 5\times10^{-6}$, $\hvec{R}: 10^{-4}$ and $\hvec{T}_\text{out}: 10^{-4}$. The number of epochs per machine learning stage is set to 1500. The number of training samples is approx.\ 300 000, chosen such that the data-to-parameter ratio equals 1. We show on Fig.\ \ref{fig:max_modes} that our characterization process reached the target $\text{TVD}_\text{test}=10^{-3}$ in 6 iterations. The number of iterations may be reduced by further tuning the learning hyperparameters (learning rates, number of epochs per machine learning stage...). This demonstrates the ability of our process to maintain data-efficiency at higher numbers of modes.

\section{PHASES COMPUTATION BY GLOBAL OPTIMIZATION}
\label{app:phase_optim}

We describe in this section the procedure for determining the phases  $\vec{\phi}$ to apply on a photonic integrated circuit (PIC) such that the implemented unitary matrix $U(\vec{\phi})$ closely approximates a target unitary matrix $V$. The cost function to minimize is

\begin{align}
    C(\vec{\phi}) 
    &= 1 - \mathcal{F}^2 \left( U(\vec{\phi}), V \right) \\
    &=1 - \left| \frac{\text{Tr} \left[ U^\dagger (\vec{\phi}) \hat{V} \right]}{m} \right|^2
\end{align}
where $\mathcal{F}$ is the fidelity metric (see App.\ \ref{app:pic_metrics}) and $m$ is the number of modes of the PIC. The cost function has a minimal value of 0 when $U(\vec{\phi})$ and $V$ differ only by a global phase factor (see App.\ \ref{app:fidelity}).  We minimize the cost function by gradient descent. The gradient of $C(\vec{\phi})$ can be expressed in closed-form as

\begin{equation}
    \frac{\partial C}{\partial \phi_j}(\vec{\phi}) = 
    \frac{2}{m^2} \Re \left\{  
    \text{Tr} \left[ U^\dagger (\vec{\phi}) \hat{V} \right]
    \text{Tr} \left[ V^\dagger \frac{\partial\hat{U}}{\partial \phi_j}(\vec{\phi}) \right].
    \right\}
\end{equation}
Following the properties of matrix-product derivation

\begin{equation}
    \frac{\partial\hat{U}}{\partial \phi_j}(\vec{\phi}) = 
    \hat{A}
    \begin{bmatrix}
    0 &   &   &  &  &  &  \\
      & \ddots &   &  &  &  &  \\
      &   & 0 &  &  &  &  \\
      &  &  & ie^{i\phi_j} &  &  &  \\
      &  &  &  & 0 &  &  \\
      &  &  &  &  & \ddots &  \\
      &  &  &  &  &  & 0
    \end{bmatrix}
    \hat{B}
\end{equation}
where $\hat{A}$ and $\hat{B}$ are the unitary matrices produced respectively by the components before and after the $j^\text{th}$ phase shifter. The optimizer is L-BFGS implemented in C++ with the {\tt NLopt} package \cite{NLopt}.

\section{MITIGATION OF INHOMOGENEOUS INPUT AND OUTPUT TRANSMISSIONS}
\label{app:losses_mitigation}

\begin{figure}
    \centering
    \includegraphics[width=\textwidth]{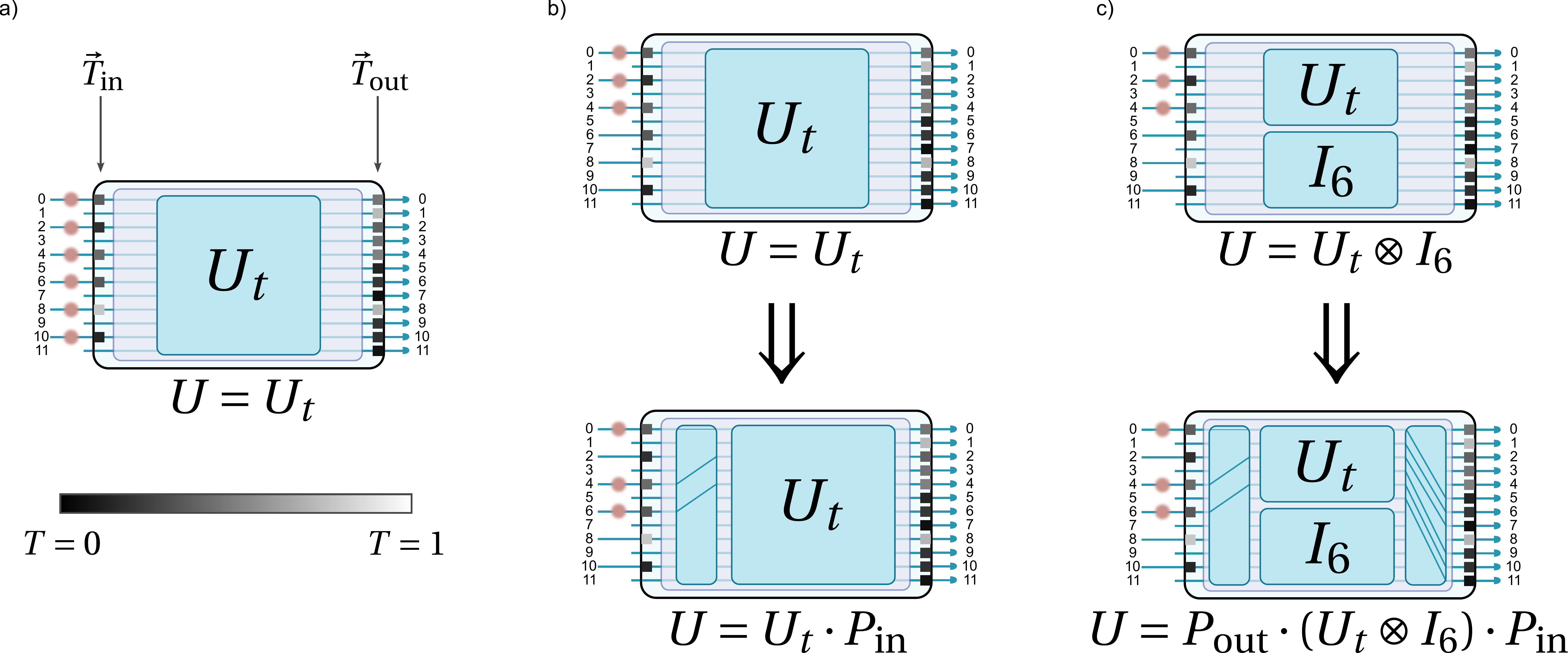}
    \caption{
    \textbf{Mitigation of inhomogeneous input and output ports transmissions for universal-scheme PICs.} 
    \textbf{a)} We consider here without loss of generality a Clements interferometer with $m=12$ input and output ports and up to $n=6$ input photons. The gray shade of input and output ports is an indicator for optical transmissions $\vec{T}_\text{in}$ and $\vec{T}_\text{out}$. In the figure, a white square corresponds to perfect transmission $T=1$. The target unitary matrix to implement on the PIC is denoted $U_t$. The pink area on the PIC corresponds to the implemented unitary matrix $U$, which is the matrix and tensor product of the blue building blocks representing smaller unitary matrices. The matrix $U$ is compiled into phases as a single unitary matrix following Sec.\ \ref{sec:compilation}, which does not limit the number, size or the depth of the building blocks. If the target matrix has dimensions $12 \times 12$ and all 6 photons are used, the target matrix $U_t$ is implemented as is.
    \textbf{b)} Mitigation techniques can be applied if the number of used photons is less than 6. Top: we choose to use 3 photons. Bottom: the imperfection mitigation selects the photons injected in the 3 input ports with the most consistent transmission values (here 0, 4 and 6 which have similar shades of gray). The input photons need hence to be routed to their respective input with a permutation $P_\text{in}$ before $U_t$. The PIC implements the unitary matrix $U=U_t \cdot P_\text{in}$.
    \textbf{c)} Input and output transmissions can also be mitigated in an analog way if the target unitary has for instance dimensions $6 \times 6$. Top:  The PIC implements here the unitary matrix $U_t \otimes I_6$ where $I_6$ is the identity matrix on 6 modes. Bottom: the imperfection mitigation selects the 3 (resp.\ 6) input ports (resp.\ output ports) with the most consistent transmissions. Then appropriate permutations $P_\text{in}$ and $P_\text{out}$ are applied before and after $U_t$. The PIC implements $P_\text{out} \cdot (U_t \otimes I_6) \cdot P_\text{in}$
    }
    \label{fig:losses_mitigation}
\end{figure}

We consider in this section universal-scheme photonic integrated circuits (PICs), which can by definition implement any unitary matrix acting on the spatial input modes. The estimated input and output transmissions $\hvec{T}_\text{in}$ and $\hvec{T}_\text{out}$ by the virtual replica from the characterization protocol (see Fig.\ \ref{fig:process} in main text) can be harnessed to mitigate imperfections. In the case of uniform losses in the PIC, i.e.\ all input ports (resp.\ output ports) have transmission $T_\text{in}$ (resp.\ $T_\text{out}$), post-selection on the number of detected photons is a type of photon loss mitigation that yields correct output probabilities. Post-selection is used in most linear optical quantum information protocols \cite{Maring2023, Mezher2023}. 

The case of inhomogeneous input and output port transmissions can be mitigated in particular cases, which we discuss in this section. We assume that the universal-scheme PIC has $m$ input and output ports, and that up to $n$ photons can be injected simultaneously into the PIC (in general $n < m$).
\begin{itemize}
    \item If the target unitary matrix $U_t$ has dimensions $m \times m$ and $n$ photons are used, the matrix is implemented as is (Fig.\ \ref{fig:losses_mitigation}a). 
    \item If the number of photons of used is strictly smaller than $n$, we select the input ports with the most homogeneous transmission values. To route the photons inside the chip, we add a permutation $P_\text{in}$ before $U_t$ (Fig.\ \ref{fig:losses_mitigation}b). Our compilation process (see Sec.\ \ref{sec:compilation}) takes as input the unitary matrix $U_t \cdot P_\text{in}$ and treats it as one single matrix to compute the corresponding phases. This mitigation technique does not impose limitations on the size of $U_t$.
    \item If the dimensions of $U_t$ are strictly smaller than $m \times m$, we select the input and output ports with the most homogeneous transmission values. Permutations $P_\text{in}$ and $P_\text{out}$ are added before and after $U_t$ (Fig.\ \ref{fig:losses_mitigation}c).
\end{itemize}

\section{PHASE-VOLTAGE RELATION SOLVER}
\label{app:solver}

We denote in this section the number of phase shifters of a photonic integrated circuit by $n_\text{PS}$. The phase-voltage relation is an equation of the form

\begin{equation}
    \vec{\phi} = \sum_{k \geq 1} C_k \vec{V}^{\odot k} + \vec{c}_0
\end{equation}
relating the implemented phases of the phase shifters $\vec{\phi}$ to the applied voltages $\vec{V}$. $\vec{c_0}$ is a vector with $n_\text{PS}$ components containing the passive phases, $^{\odot}$ represents element-wise exponentiation, and $C_k$ are $n_\text{PS} \times n_\text{PS}$ matrices. This equation possesses two additional distinctive characteristics: the off-diagonal elements of $C_k$ are very small compared to its diagonal elements, because crosstalk is expected to depend on the distance between components. In addition, the components of $\vec{\phi}$ are defined modulo $2\pi$. We devised a custom iterative solver that harnesses these features to solve the equation. 

We briefly present how the solver approximates the voltages to implement target phases $\vec{\phi}_\text{target}$. The voltages $\vec{V}$ are initialized with a zero vector. At each iteration:

\begin{enumerate}
    \item The phases $\vec{\phi}(\vec{V})$ are computed
    \item The phase difference $\vec{\phi}(\vec{V}) - \vec{\phi}_\text{target}$ is computed modulo $2\pi$.
    \item The voltages $\vec{V}$ are updated with a step proportional to the phase difference for each phase shifter.
    \item If voltage components are negative or beyond the voltage limit $V_\text{max}$, they are cast within the interval $[0, V_\text{max}]$ with a modulo operation.
\end{enumerate}

The algorithm terminates when the maximum difference between $\vec{\phi}(\vec{V})$ and $\vec{\phi}_\text{target}$ is below some specified threshold. If the algorithm remains stuck in a local minimum, a random vector is added to $\vec{V}$.

We evaluate the time complexity of the solver. The amount of iterations needed for each voltage component to converge increases linearly with $n_\text{PS}$. Evaluating $\vec{\phi}(\vec{V})$ is a matrix-vector multiplication of complexity $O(n_\text{PS}^2)$. The other operations within the iteration, like computing the difference between $\vec{\phi}(\vec{V})$ and $\vec{\phi}_\text{target}$, are $O(n_\text{PS})$. Hence, the time complexity of a single iteration is $O(n_\text{PS}^2)$. The total complexity is thus $O(n_\text{PS}^3)$.

\begin{figure}
    \centering
    \includegraphics[width=0.45\textwidth]{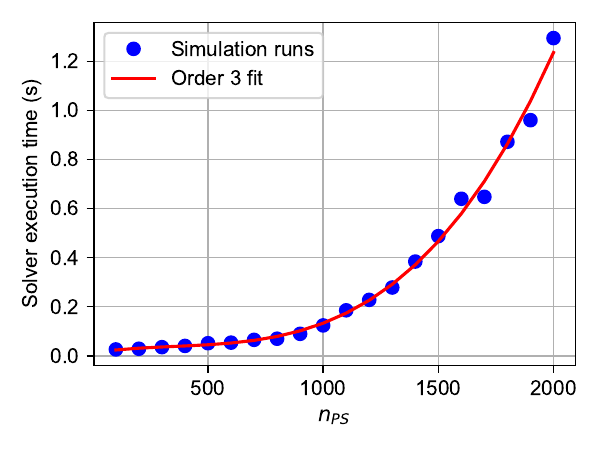}
    \caption{Phase-voltage equation solver scaling with the number of phase shifters $n_\text{PS}$}
    \label{fig:solver_scaling}
\end{figure}

We verify this by running the solver on an increasing number of phase shifters. We generate $\vec{c}_0$ with uniformly chosen values in $[-\pi, \pi]$. A $C_2$ matrix is generated by setting the diagonal elements to 0.034 and adding a gaussian random matrix with mean 0 and standard deviation 1 \% of 0.034 (this corresponds to a PIC where each phase shifter interacts with every other one with a crosstalk strength of 1 \%). We set the solver precision threshold to 0.1 mrad and $V_\text{max}=15 V$. The voltage steps are set to $\Delta\phi \times V_\text{max}/10$, where $\Delta\phi$ is the phase difference between the current and target value for a given phase shifter. The algorithm shuffles the voltages after every 500 iterations. We plot on Fig.\ \ref{fig:solver_scaling} the solver execution time against the number of phase shifters. The polynomial fit of order 3 agrees very well with the data points.

\end{document}